\input amstex  
\documentstyle{amsppt}
\magnification=\magstephalf  
 \addto\tenpoint{\baselineskip 15pt
  \abovedisplayskip18pt plus4.5pt minus9pt
  \belowdisplayskip\abovedisplayskip
  \abovedisplayshortskip0pt plus4.5pt
  \belowdisplayshortskip10.5pt plus4.5pt minus6pt}\tenpoint
\pagewidth{6.5truein} \pageheight{8.9truein}
\subheadskip\bigskipamount
\belowheadskip\bigskipamount
\aboveheadskip=3\bigskipamount
\catcode`\@=11
\def\output@{\shipout\vbox{%
 \ifrunheads@ \makeheadline \pagebody
       \else \pagebody \fi \makefootline 
 }%
 \advancepageno \ifnum\outputpenalty>-\@MM\else\dosupereject\fi}
\outer\def\subhead#1\endsubhead{\par\penaltyandskip@{-100}\subheadskip
  \noindent{\subheadfont@\ignorespaces#1\unskip\endgraf}\removelastskip
  \nobreak\medskip\noindent}
\def\endremark{\par\revert@envir\endremark\vskip\postdemoskip}
\outer\def\enddocument{\par
  \add@missing\endRefs
  \add@missing\endroster \add@missing\endproclaim
  \add@missing\enddefinition
  \add@missing\enddemo \add@missing\endremark \add@missing\endexample
 \ifmonograph@ 
 \else
 \vfill
 \nobreak
 \thetranslator@
 \count@\z@ \loop\ifnum\count@<\addresscount@\advance\count@\@ne
 \csname address\number\count@\endcsname
 \csname email\number\count@\endcsname
 \repeat
\fi
 \supereject\end}
\catcode`\@=\active
\CenteredTagsOnSplits
\NoBlackBoxes
\nologo
\def\today{\ifcase\month\or
 January\or February\or March\or April\or May\or June\or
 July\or August\or September\or October\or November\or December\fi
 \space\number\day, \number\year}
\define\({\left(}
\define\){\right)}

\define\CC{{\Bbb C}}

\define\Diff{\operatorname{Diff}}
\define\EE{\Bbb E}

\define\End{\operatorname{End}}

\define\HH{{\Bbb H}}
\define\Hom{\operatorname{Hom}}

\define\Met{\operatorname{Met}}

\define\RR{{\Bbb R}}
\define\SS{\Bbb S}

\define\Tr{\operatorname{Tr}}
\define\ZZ{{\Bbb Z}}
\define\[{\left[}
\define\]{\right]}

\define\chiup{\raise.5ex\hbox{$\chi$}}

\define\exertag #1#2{#2\ #1}

\define\index{\operatorname{index}}

\define\inv{^{-1}}

\define\protag#1 #2{#2\ #1}

\define\res#1{\negmedspace\bigm|_{#1}}
\define\temsquare{\raise3.5pt\hbox{\boxed{ }}}

\define\theprotag#1 #2{#2~#1}

\define\xca#1{\removelastskip\medskip\noindent{\smc%
#1\unskip.}\enspace\ignorespaces }

\define\zmod#1{\ZZ/#1\ZZ}

\define\zt{\zmod2}

\redefine\Re{\operatorname{Re}}

\define\rem#1{\marginalstar\begingroup\bf[{\eightpoint\smc{#1}}]\endgroup}
\def\strutdepth{\dp\strutbox} 
\def\marginalstar{\strut\vadjust{\kern-\strutdepth\specialstar}} 
\def\specialstar{\vtop to \strutdepth{ 
    \baselineskip\strutdepth 
    \vss\llap{$\bold{\Rightarrow}$ }\null}}

\NoRunningHeads 


\def\Dsl{\,\raise.15ex\hbox{/}\mkern-13.5mu D} 
\def\half{{{1\over 2}}}
\def\p{{\partial}}
\define\Ab{\overline{A}}
\define\Ad{\operatorname{Ad}}
\define\BYeff{\BY^{\operatorname{eff}}}
\define\BY{\Cal{B}_Y}
\define\CYT{\operatorname{Cliff}(\Y/T)}
\define\CY{\Cal{C}_Y}
\define\CZ {{\Cal Z}} 
\define\Cliff{\operatorname{Cliff}}
\define\C{\Cal{C}}
\define\Det{\operatorname{Det}}
\define\Dirac{D\hskip-.65em /}
\define\F{\Cal F}
\define\Hb{\overline{\H}}
\define\Hspz{\HH\spec^0}
\define\Jb{\overline{J}}
\define\Lchi{L_{\operatorname{\chi }}}
\define\Lgauge{L_{\operatorname{gauge}}}
\define\Lgrav{L_{\operatorname{gravitino}}}
\define\Lh{L^{1/2}}
\define\Ltil{\tilde{\scrL}}

\define\Pfaff{\operatorname{Pfaff}}
\define\RZ{\RR/\ZZ}
\define\SMgauge{S_{\operatorname{M2-gauge}}}
\define\Seff{S_{\operatorname{eff}}}
\define\Sgauge{S_{\operatorname{gauge}}}
\define\Sgrav{\Gamma _{\operatorname{gravitino}}}
\define\TT{\Bbb{T}}
\define\Tb{\overline{T}}
\define\Ttil{\tilde{T}}

\define\Wtil{W_S}
\define\Y{\Cal{Y}}
\define\bA{A^\partial }
\define\bC{C^\partial }

\define\bD{D^\partial }
\define\bH{\Cal{H}^\partial }
\define\bK{K^{\partial }_{\bY}}
\define\bP{P^\partial }
\define\bRS{RS^\partial }
\define\bS{S^\partial }
\define\bYY{\partial \Y}
\define\bY{\partial Y}
\define\bcY{\Cal{F}^\partial _{\bY}}
\define\bcbY{\Cal{B}^\partial _{\bY}}
\define\bc{c^\partial }
\define\bghpsi{\vphantom{M}^{\text{gh}}\!\bpsi }
\define\bg{g^\partial }
\define\bom{\omega ^{\partial }}
\define\bpsiT{\psi_T ^\partial }
\define\bpsin{\psi_\nu  ^\partial }
\define\bpsi{\psi ^\partial }
\define\btpsi{\tilde{\psi }^\partial }
\define\eBY{\overline{\Cal{B}_Y}}
\define\eBbY{\overline{\Cal{B}_{\bY}}}
\define\ev{\operatorname{ev}}
\define\ghpsi{\vphantom{M}^{\text{gh}}\!\psi }
\define\id{\operatorname{id}}
\define\pfaff{\operatorname{pfaff}}
\define\ppsi{{\psi '}}
\define\sAc{(A,c)^\sigma }

\define\sbAc{(\bA,\bc)^\sigma }

\define\scrB{\Cal{B}}
\define\scrK{\Cal{K}}
\define\scrL{\Cal{L}}

\define\sign{\operatorname{sign}}
\define\spec{\operatorname{spec}}
\define\spz{\spec^0}
\define\sqmo{\sqrt{-1}}
\define\tlambda{\tilde{\lambda }}
\define\tpsi{\tilde{\psi }}
\define\triv{\bold{1}}
\define\tr{\operatorname{tr}}
\redefine\H{\Cal{H}}

\input epsf

\define\conth#1{\medskip#1\smallskip}
\define\contsh#1#2{\par	\indent \S#1.\enspace#2\endgraf}

\refstyle{A}
\widestnumber\key{SSSSSSS}   
\document


	\topmatter
 \title\nofrills Setting the Quantum Integrand of M-Theory
\endtitle 
 \author Daniel S. Freed \\ Gregory W. Moore  \endauthor
 \thanks The work of D.F. is supported in part by NSF grant DMS-0305505.  The
work of G.M. is supported in part by DOE grant DE-FG02-96ER40949 \endthanks
 \affil Department of Mathematics, University of Texas at Austin\\ Department
of Physics, Rutgers University\endaffil
 \address Department of Mathematics, University of Texas, 1 University
Station, Austin, TX 78712-0257\endaddress 
 \email dafr\@math.utexas.edu \endemail
 \address Department of Physics, Rutgers University, Piscataway, NJ
08855-0849\endaddress 
 \email gmoore\@physics.rutgers.edu\endemail
 \date December 22, 2005\enddate
 	\abstract 
 In anomaly-free quantum field theories the integrand in the bosonic
functional integral---the exponential of the effective action after
integrating out fermions---is often defined only up to a phase without an
additional choice.  We term this choice ``setting the quantum integrand''.
In the low-energy approximation to M-theory the $E_8$-model for the $C$-field
allows us to set the quantum integrand using geometric index theory.  We
derive mathematical results of independent interest about pfaffians of Dirac
operators in $8k+3$ dimensions, both on closed manifolds and manifolds with
boundary.  These theorems are used to set the quantum integrand of M-theory
for closed manifolds and for compact manifolds with either temporal (global)
or spatial (local) boundary conditions.  In particular, we show that M-theory
makes sense on arbitrary 11-manifolds with spatial boundary, generalizing the
construction of heterotic M-theory on cylinders.
	\endabstract
	\endtopmatter

\document

The low-energy approximation to M-theory is a refinement of classical
11-dimensional supergravity.  It has a simple field content: a metric~$g$, a
$3$-form gauge potential $C$, and a gravitino.  The M-theory action contains
rather subtle ``Chern-Simons'' terms which, on a topologically nontrivial
manifold~$Y$, raise delicate issues in the definition of the (exponentiated)
action.  Some aspects of the problem were resolved by Witten~\cite{W1}.  The
key ingredients are: a quantization law for~$C$ and a background magnetic
current induced by the fourth Stiefel-Whitney class of the underlying
manifold; an expression for the exponentiated Chern-Simons terms using an
$E_8$ gauge field and an associated Dirac operator in $12$ dimensions; and
finally a sign ambiguity in the gravitino partition function.  In~\cite{DFM}
the link to~$E_8$ was used to construct a model for the $C$-field and define
precisely the action, assuming that the metric~$g$ is fixed.  The present
paper gives a complete treatment of the M-theory action as a function of both
$C$ and $g$.  Furthermore, we treat manifolds with boundary.  The boundary
may have several components and each component is interpreted either as a
fixed time slice ({\it temporal boundary\/}) or a boundary in space ({\it
spatial boundary\/}).  We do not mix temporal and spatial boundary
conditions.  Our discussion of spatial boundaries in~\S{4.3} generalizes the
case $Y= X \times [0,1]$, where $X$ is a closed $10$-manifold, which was
described in the work of Horava and Witten~\cite{HW1}, \cite{HW2}.  Our
analysis here makes it clear that the anomaly cancellation is {\it local}.
(As emphasized in~\cite{BM} the locality of anomaly cancellation in the
Horava-Witten model is far from obvious.)  In particular, we show that there
is no topological obstruction to formulating M-theory on an $11$-manifold
with an arbitrary number of boundary components, provided an independent
$E_8$ super-Yang-Mills multiplet is present on each component.

The analysis here is more than a cancellation of anomalies in M-theory.
Already in~\cite{W1} Witten showed that there is a nontrivial Green-Schwarz
mechanism canceling global anomalies on closed $11$-manifolds.  We go further
and show that the anomaly is canceled {\it canonically}.  This is a crucial
distinction for the following reason.  The absence of anomalies is a
necessary condition for a quantum theory to be well-defined, but the
cancellation mechanism depends on physically measurable choices.  Put
differently, there are undetermined phases if the configuration space of
bosonic fields is not connected.  As we explain quite generally in~\S{4.1},
the exponentiated effective action after integrating out fermionic fields is
naturally a section of a hermitian line bundle with covariant derivative over
the space of bosonic fields.  The absence of anomalies means that the line
bundle is geometrically trivializable, i.e., the covariant derivative is flat
with no holonomy.  If there are no anomalies then global trivializations
exist, and a choice of trivialization determines the integrand of the bosonic
functional integral.  When we make such a choice we say we have {\it set the
quantum integrand}.  The general uniqueness question for settings of the
quantum integrand is discussed in~\S{5.5}.  

Our main result is that in $M$-theory there is a canonical choice of
trivialization, thus a canonical setting of the quantum integrand of
$M$-theory.  The procedure by which we set the quantum integrand of
$M$-theory is, as we have mentioned, an example of the {\it Green-Schwarz
mechanism}.  Quite generally, by the Green-Schwarz mechanism we mean that
setting the quantum integrand involves a trivialization of the tensor product
of two line bundles with covariant derivative, one coming from integration
over fermionic fields and the other from the simultaneous presence of
electric and magnetic current; see~\cite{F2}, \cite{F3,Part~3} for a general
discussion.  The integral over fermionic fields is a section of a pfaffian
line bundle.  In this paper we use the $E_8$-model for the $C$-field to
define the exponentiated electric coupling.  This has the advantage that the
associated line bundle with covariant derivative is defined by
Atiyah-Patodi-Singer $\eta $-invariants associated to the $E_8$-gauge fields.
With this model, then, we can analyze both line bundles in the context of
standard invariants of geometric index theory and explicitly write down the
trivialization which sets the quantum integrand.

The mathematical results we apply to M-theory are given in~\S{1} for closed
manifolds and in~\S{3} for manifolds with boundary.  Determinant and pfaffian
line bundles are usually considered for families of Dirac operators on even
dimensional manifolds, but our interest here is in the odd dimensional case.
As we explain in~\S{1.2} there is a second natural {\it real\/} line bundle
with covariant derivative in odd dimensions, defined using the exponentiated
$\eta $-invariant, and it is isomorphic to the determinant line bundle
(\theprotag{1.16} {Proposition}).  This isomorphism is equivalent to a
trivialization of the tensor product---the trivialization needed for the
physics---since the second line bundle is real.  This isomorphism induces a
real structure on the determinant line bundle in odd dimensions.  Also, it
induces a nonflat {\it complex\/} trivialization of the determinant line
bundle, so gives a definition of the determinant of the Dirac operator in odd
dimensions as a complex number~\cite{S}; see \theprotag{1.20} {Remark}.  This
definition is often used in the physics literature, and is arrived at with
Pauli-Villars regularization~\cite{R1}, \cite{R2}, \cite{ADM}.  However, the
definition as an element of the determinant line is more fundamental.  There
is an important refinement~(\theprotag{1.31} {Proposition}) to pfaffians in
dimensions~3, 11, 19, \dots which includes the dimensions of interest in
M-theory: 11~for the bulk and 3~for M2-branes~(\S{5.2}).  This refinement is
topological in a sense made precise in Appendix~B (\theprotag{B.2}
{Proposition}).
 
We take up the generalization of this isomorphism to Dirac operators on
manifolds with boundary in section~3.  Most often considered in the geometric
index theory literature are boundary conditions of {\it global\/} type, which
in the physics correspond to a temporal boundary.  The generalization of the
basic theorem to this case is straightforward~(\S{3.2}).  {\it Local\/}
boundary conditions arise in the physics from spatial boundaries, but because
they do not exist for every Dirac operator they are less studied.  The
generalization to this case is more subtle and (in general dimensions) is the
subject of the forthcoming thesis of Matthew Scholl.  The applications to
M-theory on manifolds with boundary appear as \theprotag{4.16} {Theorem}
(temporal boundary) and \theprotag{4.35} {Conjecture} (spatial boundary).
 
Our treatment falls short by not defining precisely the partition function of
the Rarita-Schwinger (gravitino) field. The definition commonly used in the
literature seems ill-defined due to singularities related to the zeromodes of
bosonic ghosts for supersymmetry transformations.  Moreover, the derivation
of the standard expression in terms of pfaffians of Dirac operators assumes
an off-shell formulation of supergravity, something which is lacking in the
10- and 11-dimensional cases.  Nevertheless, we take the standard expression
as motivation for the line bundle with covariant derivative of which the
Rarita-Schwinger partition function is a section. We present a derivation of
the standard formula in Appendix~A, mostly to motivate the local boundary
conditions for the ghost fields which are used in section 4.3.  The precise
definition of the Rarita-Schwinger partition function is a general issue
which we leave to future work.  Another issue we do not confront is the
dependence of the covariant derivative on the Rarita-Schwinger line bundle on
background fluxes.  Nontrivial dependence can in principle arise from terms
of the form $\psi G \psi$ in the supergravity action. (There are additional
terms of a similar nature in heterotic $M$-theory.)  We believe the above
issues will not drastically alter the discussion we give, which is based on
the simple assumption that the Rarita-Schwinger partition function is a
section of the line bundle in equation~\thetag{2.2}, equipped with the
standard covariant derivative.

Some general issues of independent interest arose during our investigations.
One concerns the definition of anomalies and the setting of quantum
integrands for manifolds with temporal boundaries.  This forms part of the
discussion in~\S{4.1} and is elaborated in~\S{5.4} where we relate it to the
Hamiltonian interpretation of anomalies.  There are interesting mathematical
questions which underlie that discussion, but they are not treated here.
Another issue concerns boundary values for fields with automorphisms, such as
gauge fields.  Then the boundary condition includes a choice of isomorphism
(for example, see~\cite{FQ} where gauge theories with finite gauge groups are
treated carefully), and this shows up in the physics as certain phases, such
as $\theta $-angles.  In~\S{5.3} we indicate how this works for the $C$-field
in M-theory.
 
We thank Emanuel Diaconescu and Michael Hopkins for many discussions on these
topics and Laurent Baulieu and Edward Witten for additional discussions.  We
also thank the Kavli Institute for Theoretical Physics and Aspen Center for
Physics for providing wonderful environments in which to carry out this work.
G. M. would also like to thank the LPTHE at Jussieu, Paris for hospitality,
and the participants of the Simons Workshop in Mathematics and Physics at
SUNY Stony Brook for asking some good questions.

\bigskip\bigskip
{\eightpoint\parindent0pt
 {\bf Contents}
 \conth{Section 1: Determinants, Pfaffians, and $\eta $-Invariants}
 \contsh{1.1}{Determinant line bundle}
 \contsh{1.2}{Odd dimensions}
 \contsh{1.3}{$8k+3$ dimensions}
 \contsh{1.4}{$\zeta $-functions}
 \conth{Section 2: M-theory action on closed manifolds}
 \conth{Section 3: $(8k+3)$-Dimensional Manifolds with Boundary}
 \contsh{3.1}{Generalities}
 \contsh{3.2}{Global boundary conditions}
 \contsh{3.3}{Local boundary conditions}
 \conth{Section 4: M-Theory Action on Compact Manifolds with Boundary}
 \contsh{4.1}{Actions and Anomalies}
 \contsh{4.2}{Temporal Boundary Conditions}
 \contsh{4.3}{Spatial Boundary Conditions}
 \conth{Section 5: Further discussion}
 \contsh{5.1}{The $E_8$-model}
 \contsh{5.2}{M2-branes}
 \contsh{5.3}{Boundary values of $C$-fields}
 \contsh{5.4}{Temporal boundaries and the Hamiltonian anomaly}
 \contsh{5.5}{Topological terms}
 \conth{Appendix A: The Gravitino Path Integral}
 \contsh{A.1}{The gravitino theory} 
 \contsh{A.2}{Local analysis of the equations of motion} 
 \contsh{A.3}{Partition function} 
 \contsh{A.4}{Boundary conditions for ghosts} 
 \conth{Appendix B: Quaternionic Fredholms and Pfaffians}
}\bigskip

 \head
 \S{1} Determinants, Pfaffians, and $\eta $-Invariants
 \endhead
 \comment
 lasteqno 1@ 35
 \endcomment

The geometry of determinant line bundles was developed in~\cite{Q},
\cite{BF}; see~\cite{F1} for a survey.  In \S{1.1} we recall the main points.
Our discussion is phrased in general terms and applies in arbitrary
dimensions.  In odd dimensions Clifford multiplication by the volume form
induces a real structure on the determinant line bundle, which we explain
in~\S{1.2} by introducing a manifestly real line bundle associated to the
$\eta $-invariant~\cite{APS} and proving it is isomorphic to the determinant
line bundle.  In~\S{1.3} we prove a refinement in dimensions~$8k+3$ ($k\in
\ZZ^{\ge0}$) coming from the quaternionic structure.  Some details about
$\zeta $-functions are addressed in~\S{1.4}.

 \subhead \S1.1 Determinant line bundle
 \endsubhead

        \definition{\protag{1.1} {Definition}}
 Let $T$~be a smooth manifold.  A {\it geometric family of Dirac operators
parametrized by~$T$\/} consists of:
 \roster
 \item"\rom(i\rom)" a Riemannian manifold $\Y\to T$; that is, a fiber
bundle~$\Y\to T$, a metric on the relative tangent bundle~$T(\Y/T)\to\Y$, and
a horizontal distribution~$H$ on~$\Y$ (thus $H\oplus T(\Y/T)= T\Y$); and
 \item"\rom(ii\rom)" a bundle $M=M^0\oplus M^1\to \Y$ of complex $\zt$-graded
$\CYT$-modules with compatible metric and covariant derivative.
 \endroster  
        \enddefinition

\flushpar
 The metric and horizontal distribution determine a Levi-Civita covariant
derivative on $T(\Y/T)\to \Y$.  The Riemannian metrics determine a bundle
$\CYT\to\Y$ of (finite dimensional) Clifford algebras: the fiber at~$y\in \Y$
is the real Clifford algebra of the relative cotangent space~$T_y^*(\Y/T)$.
The Clifford module structure on~$M$ is given as a map
  $$ \gamma \:T^*(\Y/T)\longrightarrow \End(M) \tag{1.2} $$
which obeys the Clifford relation 
  $$ \gamma (\theta _1)\gamma (\theta _2) + \gamma (\theta _2)\gamma (\theta
     _1) = -2\langle \theta _1,\theta _2 \rangle,\qquad \theta _1,\theta
     _2\in T_y^*(\Y/T),\quad y\in \Y. \tag{1.3} $$
We ask that the image consist of odd skew-adjoint transformations.  The
compatibility in the last line of \theprotag{1.1} {Definition} also requires
that \thetag{1.2}~be flat.  For each~$t\in T$ the Dirac operator~$D_t =
\gamma \circ \nabla $ is defined on sections of~$M\res{\Y_t}\to\Y_t$.  It is
odd relative to the $\zt$-grading.\footnote{We remark that the $\zt$-grading
on~$M$ is {\it not\/} the physics grading of bosonic and fermionic fields.
In our exposition here sections of~$M$---for example, spinor fields---are
treated as ordinary commuting fields.  When we turn to the physics
applications in~\S{2} we use the proper action for fermionic fields.}
 
To illustrate the notation let $T$~be a point, so $D$~is a Dirac operator on
a single manifold~$Y$.  Suppose first that $\dim Y=2m$~is even and $Y$~is
spin.  Then for the standard Dirac operator $M=S$~is the bundle of
$\zt$-graded spinors with homogeneous components~$S^0,S^1$ of complex
rank~$2^{m-1}$, the bundles of chiral spinors.  The Dirac operator
interchanges the chirality of homogeneous spinor fields.  The covariant
derivative on~$S$ is induced from the Levi-Civita covariant derivative.  If
$\dim Y=2m+1$~is odd, then we usually say that spinors are ungraded: there is
no chirality.  In the $\zt$-graded setup we can take each of~$S^0,S^1$ to be
the ungraded spinor bundle of complex rank~$2^m$.  This is compatible with
the observation that for any $\zt$-graded $\Cliff(Y)$-module $M\to Y$
Clifford multiplication by the volume form provides an isomorphism~$M^0\to
M^1$ if $\dim Y$~is odd; see the next section for consequences.  For Dirac
operators with coefficients in a vector bundle~$E\to Y$ take $M=S\otimes E$.
We occasionally denote this Dirac operator as~`$D_M$'.
 
In the application to field theory the parameter space is typically an
infinite dimensional space~$\Cal{B}$ of all bosonic fields; from this point
of view we study the pullback by a map $T\to\Cal{B}$.

Given a geometric family of Dirac operators parametrized by~$T$ let 
  $$ \H=\H^0\oplus \H^1\longrightarrow T  $$
be the Hilbert space bundle whose fiber at~$t\in T$ is the space of
$L^2$~sections of~$M\res{\Y_t}\to\Y_t$.  Assume each fiber~$\Y_t$ is closed,
i.e., compact without boundary.  Then the Dirac operator $D_t$~extends to an
odd self-adjoint operator on~$\H_t$, and so~$D_t^2$ to an even self-adjoint
operator on~$\H_t$.  The spectrum of~$D_t^2$ is nonnegative, discrete, has no
accumulation points, and the eigenspaces are graded finite dimensional
subspaces of~$\H_t$.  Furthermore, if $\lambda ^2>0$~is an eigenvalue
of~$D_t^2$, then $D_t/\lambda $~is an isometry from the even component of the
$\lambda ^2$-eigenspace to the odd component.  Define $\spz(D_t^2)$ to be the
spectrum of~$D_t^2$ restricted to~$\H^0_t$.
 
There is a canonical open cover~$\{U_a\}_{a\ge0}$ of~$T$: 
  $$ U_a = \bigl\{ t\in T: a\notin \spz(D_t^2)\bigr\}. \tag{1.4} $$
On~$U_a$ we introduce the $\zt$-graded vector bundle 
  $$ \H(a) = \H^0(a)\oplus \H^1(a)\longrightarrow U_a \tag{1.5} $$
whose fiber $\H(a)_t$ at~$t\in T$ is the sum of the eigenspaces of~$D_t^2$
for eigenvalues less than~$a$.  Then $\H(a)$~is smooth of finite rank, with
constant rank on each component of~$U_a$.  Furthermore, the geometric data
induces a metric and covariant derivative on~$\H(a)$.  The Dirac operator~$D$
restricts to an operator~$D(a)$ on~$\H(a)$.  Global geometric invariants of
Dirac operators are constructed by patching invariants on~$U_a$.

Recall that the determinant line~$\Det\EE$ of a finite dimensional ungraded
vector space~$\EE$ is its highest exterior power.  A linear map
$S\:\EE^0\to\EE^1$ between vector spaces of the same dimension has a
determinant 
  $$ \det S\;\in\; \Hom(\Det\EE^0,\Det\EE^1)\cong \Det\EE^1\otimes
     (\Det\EE^0)^* \tag{1.6} $$
which is the induced map on the highest exterior power.  The line which
appears in~\thetag{1.6} is the determinant line of the $\zt$-graded vector
space~$\EE^0\oplus \EE^1$.  It is natural to grade~$\Det\EE$ by~$\dim\EE$.
For our purposes we take the grading to lie in~$\zt$ rather than~$\ZZ$.

Returning to the family of Dirac operators, define the line bundle
  $$ \Det\H(a) = \Det \H^1(a) \otimes \Det\H^0(a)^*\longrightarrow
     U_a.  $$
For~$b>a$ we set
  $$ \H(a,b) = \H^0(a,b)\oplus \H^1(a,b)\longrightarrow U_a\cap U_b
      $$
whose fiber at~$t$ is the sum of the eigenspaces of~$D_t^2$ for eigenvalues
between~$a$ and~$b$.  There is a canonical isomorphism 
  $$ \Det \H(a) \otimes \Det\H(a,b)\longrightarrow \Det\H(b)\qquad \text{on
     $U_a\cap U_b$} \tag{1.7} $$
and a canonical nonzero section\footnote{For ease of notation we write `$\det
D$' instead of the more accurate~`$\det D^0$'.}~$\det D(a,b)$ of~$\Det
\H(a,b)$, where $D(a,b)_t\:\H^0(a,b)_t\to \H^1(a,b)_t$ is the restriction of
the ``chiral'' Dirac operator~$D^0_t\:\H_t^0\to\H^1_t$.  From~\thetag{1.7}
we obtain the patching isomorphism
  $$ \Det\H(a)\longrightarrow \Det\H(b)\qquad \text{on $U_a\cap U_b$},
     \tag{1.8} $$
and a cocycle identity on~$U_a\cap U_b\cap U_c$, whence a global smooth line
bundle $\Det D\to T$.  Furthermore, the sections~$\det D(a)$ of~$\Det\H(a)$,
defined analogously to~$\det D(a,b)$, patch to a smooth section~$\det D$ of
$\Det D\to T$.
 
The patching isomorphism~\thetag{1.8} preserves the $\zt$ grading of the
determinant line: the parity of~$\Det D_t$ is the parity of~$\index D^0_t$.
 
The metric and covariant derivative on~$\H(a)$ induce a metric and covariant
derivative on~$\Det\H(a)$, but these are not preserved by ~\thetag{1.8}.
Modify the metric and covariant derivative to obtain invariance under
patching: multiply the metric on~$\Det \H(a)_t$ by
  $$ \prod\limits_{{a<\lambda ^2}\atop{\lambda ^2\in \spz(D_t^2)}} \lambda ^2
     \tag{1.9} $$
and add the 1-form 
  $$ \Tr\bigl(\nabla D\circ D\inv \res{\H^0 \ominus \H^0(a)} \bigr)
     \tag{1.10} $$
to the covariant derivative on~$\Det\H(a)$.  We use $\zeta $-function
regularization to define~\thetag{1.9} and~\thetag{1.10}; see~\S{1.4} for
details.

        \proclaim{\protag{1.11} {Proposition}}
 Let $D$~be a geometric family of Dirac operators on closed manifolds
parametrized by~$T$.  Then there is a functorially associated $\zt$-graded
complex line bundle
  $$ \Det D\longrightarrow T  $$
with metric and covariant derivative, and a section~$\det D$ which vanishes
at~$t\in T$ for which $D_t$~has a nonzero kernel. 
        \endproclaim

\flushpar
 There are formulas for the holonomy and curvature of the determinant line
bundle, but we will not need them in this paper.

 \subhead \S1.2 Odd dimensions
 \endsubhead

Now suppose the fibers of $\Y\to T$ have odd dimension and are oriented.  Let
$\omega $~denote Clifford multiplication by the relative volume form; it is
an odd endomorphism of~$M$.  Multiply by a suitable power of~$\sqrt{-1}$ to
arrange~$\omega ^2=1$.  Also, $\omega $~commutes with~$\gamma (\theta )$ for
any relative cotangent vector~$\theta $ and is flat, so commutes with the
Dirac operator~$D$ on any fiber.  (This commutation is {\it not\/} in the
graded sense.)  The composition~$\omega D$ is a first-order even self-adjoint
operator with discrete real spectrum, the spectrum has no accumulation
points, and the spectrum is unbounded both positively and negatively.
Observe $(\omega D)^2=D^2$ so that eigenvalues of~$\omega D$ square to
eigenvalues of~$D^2$.  The self-adjoint operator~$\omega D$ on~$\H^0$ is what
is usually termed the Dirac operator in odd dimensions.  Let $\spz(\omega
_tD_t)$ denote the spectrum of~$\omega _tD_t$ on~$\H^0_t$ for each~$t\in T$;
an eigenvalue is repeated in~$\spz(\omega _tD_t)$ according to its
multiplicity.
 
Define the open cover~$\{V_\alpha \}_{\alpha \in \RR}$ of~$T$:
  $$ V_\alpha = \bigl\{ t\in T: \alpha \notin \spz(\omega _tD_t)\bigr\}.
      $$
Note that $U_{\alpha ^2} = V_\alpha \cap V_{-\alpha }$.  Define 
  $$ \eta (\alpha )\:V_\alpha \to\RR $$
to be the $\zeta $-function regularization of $ \#\!\left\{ {\lambda \in
\spz(\omega _tD_t)}: {\alpha <\lambda }\right\} - \#\!\left\{ {\lambda \in
\spz(\omega _tD_t)} : {\lambda <\alpha }\right\}$.  Namely, for $s\in \CC$
with $\Re s>>0$ define
  $$ \eta (\alpha )_t[s]=\sum\limits_{\lambda \in
     \spz(\omega_tD_t)\;\setminus\; \{0\} } \sign(\lambda -\alpha )|\lambda
     |^{-s} \;-\; \sign(\alpha )\cdot \#\left\{ \spz(\omega_tD_t)\;\cap\;
     \{0\} \right\}
       $$
and set $\eta (\alpha )_t$ to be the value of the meromorphic continuation of
$\eta (\alpha )_t[s]$ at ${s=0}$.  For~$\alpha <\beta $ we have
  $$ \frac{\eta (\beta )_t}{2} = \frac{\eta (\alpha )_t}{2} \;-\; \#\left\{
     {\lambda \in \spz(\omega _tD_t)} :{\alpha <\lambda <\beta }\right\} \qquad
     \text{on $V_\alpha \cap V_\beta $}. \tag{1.12} $$
We use~\thetag{1.12} to construct two invariants.  First, let $\TT$~denote
the group of unit norm complex numbers.  Then
  $$ \tau (\alpha ) = \exp\left( 2\pi i\;\frac{\eta (\alpha )}{2}
     \right)\:V_\alpha \longrightarrow \TT \tag{1.13} $$
is invariant under patching, so defines a global function 
  $$ \tau_D \:T\longrightarrow \TT. \tag{1.14} $$
Second, we use the integers in the last term of~\thetag{1.12} to patch a
principal $\ZZ$-bundle on~$T$: the fiber at~$t\in T$ is 
  $$ \left\{n\:\RR\setminus \spz(\omega _tD_t) \longrightarrow \ZZ \;:\;
     n(\beta ) \;=\; n(\alpha )\,-\,\#\left\{\tsize{{\alpha <\lambda<\beta
     }\atop{\lambda \in \spz(\omega
     _tD_t)}}\right\}\vphantom{a\res{\res\beta}} \right\}.  $$
The spectral flow of~$\omega D$ around a loop in~$T$ is the monodromy of this
principal $\ZZ$-bundle.  Also, $\eta /2$~is a section of the associated real
affine bundle constructed from the translation action of~$\ZZ$ on~$\RR$.
Topologically, each construction determines a class in~$H^1(T;\ZZ)$, and the
classes are equal.  The reduction to~$H^1(T;\zt)$ is represented
geometrically by a complex line bundle~$L\to T$ with compatible real
structure, metric, and covariant derivative.  Its fiber at~$t\in T$ is
  $$ L_t = \left\{f\:\RR\setminus \spz(\omega _tD_t) \longrightarrow \CC
     \;:\; f(\beta ) \;=\; (-1)^{\#\left\{{{\alpha <\lambda<\beta
     }\atop{\lambda \in \spz(\omega _tD_t)}}\right\}}\;f(\alpha )
     \right\}. \tag{1.15} $$
Over~$V_\alpha $ the map $f\mapsto f(\alpha )$ gives an isomorphism with the
trivial bundle, and we use it to define the real structure, metric, and
covariant derivative.  Note that the covariant derivative has order two, that
is, $L^{\otimes 2}$~is canonically geometrically trivial.

On $U_a\subset T$ the Clifford multiplication~$\omega $ restricts to a flat
isometry $\omega (a)\:\H^0(a)\to\H^1(a)$, so induces a
trivialization~$\det\omega $ of~$\Det\H(a)\to U_a$.  We use it to identify
the determinant line bundle with the line bundle~$L$.

        \proclaim{\protag{1.16} {Proposition}}
  Let $D$~be a geometric family of Dirac operators on closed odd dimensional
manifolds parametrized by~$T$.  Then there is a functorial
trivialization~$\triv$ of $L\otimes\Det D\to T$ which is geometric in the
sense that
  $$ \aligned
      |\triv| &= 1 \\
      \nabla \triv&=0.\endaligned \tag{1.17} $$ 
It induces a real structure on~$\Det D$ with respect to which the
section~$\det D$ is real.
        \endproclaim

        \demo{Proof}
The fiber of~$L \otimes\Det D$ at~$t\in T$ is
  $$ (L  \otimes \Det D)_t = \left\{g\:\RR\setminus \spz(\omega _tD_t)
     \longrightarrow \Det D_t \;:\; g(\beta ) \;=\; (-1)^{\#\left\{{{\alpha
     <\lambda<\beta }\atop{\lambda \in \spz(\omega _tD_t)}}\right\}}
     \;g(\alpha ) \right\}. \tag{1.18} $$
Let~$\alpha \ge0$.  Define~$\triv$ on~$U_{\alpha ^2}$ by
  $$ g(-\alpha ) = \frac{\det\omega (\alpha ^2)}{\left|\det \omega
     D\res{\H^0\ominus \H^0(\alpha ^2)}\right|}\quad \in\quad \Det \H(\alpha
     ^2). \tag{1.19} $$
If~$\beta >\alpha $ then under the patching~\thetag{1.8} we have 
  $$ \spreadlines{5pt}\split
      g(-\alpha ) \longmapsto \quad &\frac{\det\omega (\alpha ^2)\det
     D(\alpha ^2,\beta ^2)}{\left|\det\omega D\res{\H^0\ominus\H^0(\alpha
     ^2)}\right|} \\
      =\quad &\frac{\det\omega (\alpha ^2)\det \omega (\alpha ^2,\beta
     ^2)\det \omega D(\alpha ^2,\beta ^2)}{\left|\det\omega
     D\res{\H^0\ominus\H^0(\alpha ^2)}\right|} \\
      =\quad &(-1)^{\#\left\{{{-\beta <\lambda<-\alpha }\atop{\lambda \in
     \spz(\omega _tD_t)}}\right\}}\;\frac{\det\omega (\beta
     ^2)}{\left|\det\omega D\res{\H^0\ominus\H^0(\beta ^2)}\right|} \\
      =\quad &(-1)^{\#\left\{{{-\beta <\lambda<-\alpha }\atop{\lambda \in
     \spz(\omega _tD_t)}}\right\}}\;g(-\beta ) \endsplit  $$
which matches~\thetag{1.18}.  Thus~$\triv$ is well-defined.  The
definition~\thetag{1.9} of the metric shows~$|\triv|=1$.  We verify $\nabla
\triv=0$ at the end of~\S{1.4}.  Finally, $\triv$~induces an isomorphism
$\Det D\cong L$, in view of the isomorphism~$L\cong L\inv $, and if $\det
D_t\not= 0$ then $\det D_t$~corresponds to the function~$f\in L_t$ whose
value at~0 is $f(0)=|\det \omega _tD_t|$ ($\zeta $-regularized), so is real.
        \enddemo

        \remark{\protag{1.20} {Remark}}
 The real structure on~$\Det D$ is a bit indirect, as opposed, say, to the
real structure on the pfaffian line bundle in $8k+3$~dimensions, which is
defined in the next subsection directly in terms of the geometry.  In fact,
there is a natural {\it complex\/} trivialization of~$L$, namely the square
root~$\tau_D ^{1/2}$ of~\thetag{1.14}.  In the notation of~\thetag{1.15} it
is defined by $f(\alpha )= \exp\bigl(2\pi i\eta (\alpha )/4 \bigr)$.  Then
\theprotag{1.16} {Proposition} renders the ratio $\det D\cdot \tau_D ^{-1/2}$
a global complex function.  It is sometimes defined to be the determinant in
odd dimensions, both in the mathematics~\cite{S,\S4} and physics
literature~\cite{R1}, \cite{R2}, \cite{ADM}.  Notice that the
trivialization~$\tau_D ^{1/2}$ has unit norm but is {\it not\/} flat.
Rather, we can modify the covariant derivative on~$L$ by the imaginary
1-form~$-\nabla \tau_D ^{1/2}$ and then $\tau_D ^{1/2}$ is flat relative to
the new covariant derivative.  The 1-form used to modify the covariant
derivative has a local formula---up to a factor it is the 1-form component of
the integral of the usual index density over the fibers of~$\Y\to T$.
        \endremark

     \subhead \S1.3 $8k+3$ dimensions  
     \endsubhead

Assume the fibers of $\Y\to T$ are closed of dimension~$8k+3$ for some
integer~$k\ge 0$.  Then the fibers of $\Cliff(\Y/T)\to\Y$ have a quaternionic
structure.\footnote{The complex conjugate~$\overline{\EE}$ of a complex
vector space~$\EE$ is defined thus: $\overline{\EE}=\EE$ as sets, the
additional laws on~$\EE$ and~$\overline{\EE}$ are equal, the scalar
multiplication is conjugated.  Write~$\bar{e}\in \overline{\EE}$ for the
element which equals~$e\in \EE$.  Then for~$\lambda \in \CC$ we have
$\bar{\lambda }\cdot \bar{e}=\overline{\lambda \cdot e}$.  A quaternionic
structure is a {\it linear\/} map $J\:\EE\to\overline{\EE}$ such
that~$\overline{J}J=-\operatorname{id}_{\EE}$.  (A linear map
$\EE\to\overline{\EE}$ is often termed an antilinear map on~$\EE$, but we
prefer to use only linear maps.)}  Thus assume also that the fibers
of~$M\to\Y$ are quaternionic\footnote{In the physics literature the
quaternionic structure on spinor fields is usually written ``$J\psi = C
\psi^*$'', where $C$ is the charge conjugation matrix.  Hence $J$~is usually
regarded as anti-linear.}  and that the geometry respects the quaternionic
structure.\footnote{To wit, if $J\:M\to \overline{M}$ denotes the
quaternionic structure, then $J$~is unitary and $\nabla J=0$.  The unitarity
implies
  $$ \langle J\psi ,\ppsi \rangle = -\langle J\ppsi,\psi
     \rangle,\qquad \psi ,\ppsi\in M.  $$
In addition $\gamma (\theta )\in \End M$ is quaternion linear for each
cotangent vector~$\theta $.}  For example, if the fibers of~$\Y\to T$ are
spin, then we can take $M$~to be the relative spin bundle, or the relative
spin bundle tensored with a {\it real\/} vector bundle on~$\Y$.  In this case
the determinant and $\eta $-invariant have refinements due to the fact that
the eigenspaces of~$D_t^2$ are quaternionic, so have even complex dimension.

Let $J\: M\to \overline{M}$ be the quaternionic structure.  Then $J$~commutes
with~$\omega $ and~$D$, hence for each~$t\in T$ with 
  $$ \Dirac_t=J\omega _tD_t\:\Gamma _{\Y_t}(M^0)\longrightarrow
     \Gamma_{\Y_t}(\overline{M^0}), \tag{1.21} $$
and the latter is formally skew-adjoint:
  $$ \int_{\Y_t}\bigl\{ \langle {\Dirac_t\psi _1},\psi _2 \rangle +
     \langle \Dirac_t\psi _2 ,{\psi _1}\rangle \bigr\}\;|dy| = 0
     \tag{1.22} $$
for all sections~$\psi _1,\psi _2\in \Gamma _{\Y_t}(M^0)$
of~$M^0\res{\Y_t}\to\Y_t$.  Here $\langle \cdot ,\cdot \rangle$~is the
hermitian metric on~$M$---a homomorphism from $\overline{M}\otimes M$ to the
trivial bundle---and $|dy|$~is the Riemannian measure on~$\Y_t$.  The
operator~\thetag{1.21} extends to a skew-adjoint operator
$\Dirac_t\:\H_t^0\to \overline{\H}^0_t\cong (\H_t^0)^*$.
 
A skew-adjoint operator $S\:\EE\to \EE^*$ on a finite dimensional complex
vector space~$\EE$ is equivalently a 2-form~$\omega _S\in {\tsize\bigwedge}
^2\EE^*$.  Suppose $\dim\EE=2n$~is even.  Then the {\it pfaffian\/} of~$S$ is 
  $$ \pfaff S=\frac{\omega _S^n}{n!}\in \Det\EE^* \tag{1.23} $$
and the graded line~$\Det\EE^*$ has even parity.  If $\dim \EE$~is odd, then
$\pfaff S=0$ and the parity of~$\Det\EE^*$ is odd.  In all cases $(\pfaff
S)^{\otimes 2} =\det S$ as elements of~$(\Det\EE^*)^{\otimes 2}$.  If
$J\:\EE\to\overline{\EE}$ is a quaternionic structure, then of course $\dim
\EE$ ~is even.  Furthermore, $\det J\:\Det\EE\to\overline{\Det\EE}$ is a real
structure.  If $S$~commutes with~$J$, then $\det J(\pfaff S) =
\overline{\pfaff S}$, i.e., $\pfaff S$~is a real element of~$\Det\EE^*$.
 
Recall from~\thetag{1.4}, \thetag{1.5} the open cover~$\{U_a\}_{a\ge0}$
of~$T$ and the finite rank complex vector bundles $\H^0(a)\to U_a$.  Now
$\Dirac$~restricts to a skew-adjoint operator
$\Dirac(a)\:\H^0(a)\to\H^0(a)^*$, so defines 
  $$ \pfaff\Dirac(a)\in \Det\H^0(a)^*.  $$
Since $\H^0(a)$~is quaternionic, $\Det\H^0(a)^*$~has even parity, a real
structure, and $\pfaff\Dirac(a)$~is even.  Analogous to~\thetag{1.8} we patch
on~$U_a\cap U_b$ using $\pfaff\Dirac(a,b)\in \Det\H^0(a,b)^*$.  To patch the
metric and covariant derivative we alter the correction factors~\thetag{1.9}
and~\thetag{1.10} using the fact that every eigenvalue has even
multiplicity.  Namely, write $\spz(D_t^2) = \Hspz(D_t^2) \cup \Hspz(D_t^2)$ and
replace~\thetag{1.9} by its square root
  $$ \prod\limits_{{a<\lambda ^2}\atop{\lambda ^2\in \Hspz(D_t^2)}} \lambda ^2
      \tag{1.24} $$
and~\thetag{1.10} by 
  $$ \Tr_{\HH}\bigl(\nabla D\circ D\inv \res{\H^0 \ominus \H^0(a)} \bigr)
      \tag{1.25} $$
Here `$\Tr_{\HH}$'~is the trace of a quaternion linear operator on a
quaternionic vector space.

        \proclaim{\protag{1.26} {Proposition}}
 Let $D$~be a geometric family of Dirac operators on closed
$(8k+3)$-dimensional manifolds parametrized by~$T$.  Then there is a
functorially associated complex line bundle
  $$ \Pfaff \Dirac\longrightarrow T  $$
of even parity with metric and covariant derivative, a compatible real
structure, and a real section~$\pfaff \Dirac$ which vanishes at~$t\in T$ for
which $D_t$~has a nonzero kernel.  There is a canonical isomorphism $\Det
D\cong (\Pfaff \Dirac)^{\otimes 2}$ which preserves the real structure,
metric, and covariant derivative.  Under this isomorphism $\det D = (\pfaff
\Dirac)^{\otimes 2}$.
        \endproclaim

        \remark{\protag{1.27} {Remark}}
 Since the covariant derivative on~$\Pfaff \Dirac$ has order two, it follows
that $\Det D$~has a canonical geometric trivialization, i.e., a
section~$\triv$ which satisfies~\thetag{1.17}.
        \endremark

        \demo{Proof}
 We comment only on the isomorphism.  On~$U_a$ it is given as 
  $$ \det(J\omega )\otimes \operatorname{id_{\Det \H^0(a)^*}}:\Det
     \H^1(a)\otimes \Det\H^0(a)^*\longrightarrow (\Det \H^0(a)^*)^{\otimes
     2}.  $$
The isomorphism commutes with the patching~\thetag{1.8} on~$U_a\cap U_b$ and
carries~$\det D$ to~$(\pfaff\Dirac)^{\otimes 2}$.  The latter fact shows that
it preserves the real structures; that it preserves the metrics and covariant
derivatives follows from the formulas \thetag{1.9}/\thetag{1.24} and
\thetag{1.10}/\thetag{1.25}.
        \enddemo

The quaternionic structure also leads to a refinement of the $\eta
$-invariant.  

        \proclaim{\protag{1.28} {Proposition}}
 Let $D$~be a geometric family of Dirac operators on closed
$(8k+3)$-dimensional manifolds parametrized by~$T$.  Then there is a
functorially associated global function
  $$ \tau_D ^{1/2}\:T\longrightarrow \TT \tag{1.29} $$
whose square is~\thetag{1.14}. 
        \endproclaim

\flushpar
 To construct~$\tau_D ^{1/2}$ we simply modify~\thetag{1.12} to
  $$ \frac{\eta (\beta )_t}{4} = \frac{\eta (\alpha )_t}{4} \;-\; \#\left\{
     {\lambda \in \Hspz(\omega _tD_t)} :{\alpha <\lambda <\beta }\right\}
     \qquad \text{on $V_\alpha \cap V_\beta $}  $$
and exponentiate as in~\thetag{1.13}.  The quaternionic structure also leads
to a principal $\ZZ$-bundle over~$T$ whose square (or double) is the one
mentioned after~\thetag{1.14}, as well as a square root of the line
bundle~$L\to T$ constructed in~\thetag{1.15}:
  $$ L^{1/2}_t = \left\{f\:\RR\setminus \Hspz(\omega _tD_t) \longrightarrow \CC
     \;:\; f(\beta ) \;=\; (-1)^{\#\left\{{{\alpha <\lambda<\beta
     }\atop{\lambda \in \Hspz(\omega _tD_t)}}\right\}}\;f(\alpha ) \right\}
     . \tag{1.30} $$
Note that the square $L\to T$ has a canonical geometric trivialization.

Finally, we have the following refinement of \theprotag{1.16} {Proposition}.

        \proclaim{\protag{1.31} {Proposition}}
  Let $D$~be a geometric family of Dirac operators on closed
$(8k+3)$-dimensional manifolds parametrized by~$T$.  There is a functorial
geometric trivialization~$\triv$ of $L^{1/2}\otimes\Pfaff \Dirac$.
        \endproclaim

\flushpar
 The trivialization~$\triv$ is real, in addition to satisfying~\thetag{1.17}.
\theprotag{1.31} {Proposition} is essentially a topological statement, as
is explained in Appendix~B.

        \demo{Proof}
 Replace~\thetag{1.19} with 
  $$ f(-\alpha ) = \frac{\pfaff J(\alpha ^2)}{\left|\pfaff \omega
     D\res{\H^0\ominus \H^0(\alpha ^2)}\right|}\quad \in\quad \Det
     \H^0(\alpha ^2)^*,  $$
where $J(\alpha ^2)$~is the restriction of~$J$ to~$\H^0(\alpha ^2)$ and the
denominator is the $\zeta $-regularized product
  $$ \prod\limits_{{\alpha <|\lambda |}\atop{\lambda \in \Hspz(\omega
     _tD_t)}} |\lambda |.  $$
        \enddemo

        \remark{\protag{1.32} {Remark}}
 The square root~$\tau_D ^{1/4}$ of~\thetag{1.29} is a {\it complex\/}
section of~$L^{1/2}$.  In the notation of~\thetag{1.30} it is defined by
$f(\alpha )= \exp\bigl(2\pi i\eta (\alpha )/8 \bigr)$.  Its role is not
analogous to that of~$\tau_D ^{1/2}$ in \theprotag{1.20} {Remark}, which is
used as a trivialization to define a complex determinant: the real structure
on~$\Pfaff \Dirac$ comes directly from the geometry and hence should be
respected.     
 
A direct consequence of \theprotag{1.31} {Proposition} is that the product
  $$ \Pfaff\Dirac\cdot \tau _D^{1/4}\:T\longrightarrow \CC  $$
is a well-defined global function.  This product appears as part of the
M-theory action; see~\S{2}.
        \endremark

 \subhead \S1.4 $\zeta $-functions
 \endsubhead

Let $D$~be a geometric family of Dirac operators on closed manifolds
parametrized by~$T$, and work on the open set~$U_a\subset T$ of
~\thetag{1.4}.  Define the $\zeta $-functions 
  $$ \zeta (a)[s] = \Tr\left( (D^2)^{-s}\res{\H^0\ominus\H^0(a)} \right)
     \tag{1.33} $$
and 
  $$ A(a)[s] = \Tr\left( (D^2)^{-s}\nabla D\,D\inv \res{\H^0\ominus\H^0(a)}
     \right) .  $$
For $\Re(s)$ sufficiently large $\zeta (a)[s]$~is a smooth function and
$A(a)[s]$~a smooth 1-form on~$U_a$.  The basic analytic results are: the
function~ $\zeta (a)[s]$~has a meromorphic continuation to~$s\in \CC$ which
is holomorphic at~$s=0$; the function~ $A(a)[s]$~has a meromorphic
continuation to~$s\in \CC$ with a simple pole at~$s=0$.  The product
in~\thetag{1.9} is, by definition,
  $$ \exp\left( -\frac{d}{ds}\negmedspace\Bigm|_{s=0}\zeta
     (a)[s]\right) $$
and the trace in~\thetag{1.10} is 
  $$ \frac{d}{ds}\negmedspace\Bigm|_{s=0}\Bigl( s\,A(a)[s]
     \Bigr).  $$
Rewrite~\thetag{1.33} as 
  $$ \zeta (a)[s] = \Tr\left( (D^1D^0)^{-s} \right),  $$
where $D=\left(\smallmatrix 0&D^1\\D^0&0  \endsmallmatrix\right)$ relative to
$\H=\H^0\oplus \H^1$ and we omit from the notation the restriction to
$\H^0\ominus \H^0(a)$.  Applying the differential on~$T$ we find 
  $$ d\zeta (a)[s] \,=\, -s\Tr\left( (D^1D^0)^{-(s+1)}(\nabla D^1\circ D^0 +
     D^1\circ \nabla D^0) \right).  \tag{1.34} $$
Since $D$~is self-adjoint, and using the cyclicity of the trace, we conclude 
  $$ d\zeta (a)[s] \,=\, -2s\Re\Tr\left( (D^1D^0)^{-s}\nabla D^1\circ
     (D^1)\inv \right) = -2s\Re A(a)[s]. \tag{1.35} $$
These manipulations are valid for~$\Re s>>0$, and by analytic continuation
for all~$s$.  The $s$-derivative at~$s=0$ is used in the proof of
\theprotag{1.11} {Proposition} to show that the metric and covariant
derivative on~$\Det D$ are compatible.
 
Now assume the manifolds are odd dimensional, as in~\S{1.2}.  Then $\omega
D^0=D^1\omega $ and we have 
  $$ \split
      \Tr\left( (D^1D^0)^{-(s+1)}D^1\nabla D^0 \right)
       &= \Tr\left( (D^1D^0)^{-(s+1)}D^1\omega ^2\nabla D^0 \right) \\
       &= \Tr\left( (D^1D^0)^{-(s+1)}\omega D^0\nabla D^1\omega \right) \\
       &= \Tr\left(\omega (D^0D^1)^{-s} (D^1)\inv \nabla D^1\omega \right) \\
       &= \Tr\left(\omega D^0 (D^1D^0)^{-(s+1)} \nabla D^1\omega \right) \\
       &= \Tr\left((D^1D^0)^{-(s+1)} \nabla D^1\circ D^0 \right)\endsplit
      $$
Comparing with~\thetag{1.34} and~\thetag{1.35} we see that $A(a)[s]$~is real,
and so we may omit~`$\Re$' in~\thetag{1.35}. 
 
Now we complete the proof of \theprotag{1.16} {Proposition} by checking
that~$\nabla \triv=0$, where $\triv$, defined in~\thetag{1.19}, is in our
current notation 
  $$ g(-\alpha ) = \frac{\det\omega (\alpha ^2)}{\exp\left( -\frac
     12\frac{d}{ds}\negmedspace\Bigm|_{s=0}\zeta (\alpha ^2)[s]\right)} \quad
     \in\quad \Det \H(\alpha ^2).  $$
The covariant derivative on~$\Det\H(\alpha ^2)$ is $\nabla ^0 +
\frac{d}{ds}\negmedspace\Bigm|_{s=0}\negmedspace\!(sA(\alpha
^2)[s])$ for $\nabla ^0$~the
natural covariant derivative.  Note~$\nabla ^0\det\omega (\alpha ^2)=0$.  A
short computation shows that \thetag{1.35} (without~`$\Re$') immediately
implies~$\nabla \triv=0$.

 \head
 \S{2} M-theory action on closed manifolds
 \endhead
 \comment
 lasteqno 2@ 10
 \endcomment

M-theory is an 11-dimensional theory.  Let $Y$~be an 11-manifold, which in
this section we assume is closed.  There are two bosonic fields: a Riemannian
metric~$g$ and a field~$C$ which is locally a 3-form on~$Y$.  We review the
$C$-field below.  There is a single fermionic field, the gravitino or
Rarita-Schwinger field~$\psi $, and to accommodate it we also assume that
$Y$~is spin.\footnote{M-theory respects parity reversal, so exists on
non-oriented and even non-orientable manifolds (which then carry a
certain pin structure).  In this paper, though, we assume that all manifolds
are oriented.}  Let $S^0\oplus S^1\to Y$ be the basic $\zt$-graded complex
$\Cliff(Y)$-module.  As explained in the paragraph following~\thetag{1.3} we
may take~$S^0=S^1$ to be the standard ungraded rank~32 complex spin bundle.
Set~$RS=RS^0\oplus RS^1=S\otimes T^*Y$.  Then the Rarita-Schwinger
field~$\psi $ is a section of~$RS^0$.  Note that~$S$, and so also~$RS$,
carries a quaternionic structure.  Let $D_S$~be the Dirac operator on~$S$ and
$D_{RS}$~the Dirac operator on~$RS$.   
 
The effective action for the gravitino is usually written as
  $$ \exp(-\Sgrav) = ``\vphantom{\Biggl(}\frac{\pfaff \Dirac_{RS}}{(\pfaff
     \Dirac_S)^3}\text{''}, \tag{2.1} $$
which is a function of the metric~$g$.  See Appendix~A for a formal
derivation and discussion.  Recall also from the introduction that our
treatment drops certain terms from the supergravity action.  While
\thetag{2.1}~is problematic, what seems certain is that in a definitive
treatment of the gravitino the effective action is a section of the line
bundle
  $$ \Pfaff \Dirac_{RS} \otimes (\Pfaff \Dirac_S)^{\otimes
     (-3)}\longrightarrow T \tag{2.2} $$
over any family of Riemannian spin manifolds $\Y\to T$.  In~\S{1.3} we showed
that \thetag{2.2}~carries a natural real structure, metric, and covariant
derivative.  Furthermore, its square is canonically trivial (including its
geometry).
 
We use the model for the $C$-field expounded in detail in~\cite{DFM}, and
defer to that paper for details.  A $C$-field is an object in a
groupoid~$\CY$.  In the quantum theory one is instructed to integrate over
the space of equivalence classes.  The groupoid~$\CY$ depends on the
Riemannian metric~$g$, so the space of equivalence classes of bosonic fields
in M-theory is a fiber bundle: the base is the space of equivalence classes
of metrics, fiber the space of equivalence classes of $C$-fields for a fixed
metric.  An object in~$\CY$ is a pair~$(A,c)$ consisting of a connection~$A$
on a principal $E_8$-bundle\footnote{In the notation `$A$'~is understood to
signify the connection as well as the bundle~$P$ which carries it.}~$P\to Y$
and a 3-form~$c\in \Omega ^3(Y)$.  We do not review in detail the morphisms
or the space of equivalence classes; see~\cite{DFM,\S3}.  Our interest here
is in the factor in the exponentiated action, which is usually written in
terms of a local 3-form~$C$ as
  $$ ``\vphantom{\Biggl(}\exp\left( 2\pi i \int_{Y}\Bigl\{\frac 16 C\wedge
     G\wedge G - C\wedge I_8(g) \Bigr\} \right)\text{''}, \tag{2.3} $$
where $I_8(g)$~is a certain combination of Pontrjagin forms (representing
$(4p_2-p_1^2)/192$).

We now make~\thetag{2.3} precise in this model for the $C$-field.
 
        \definition{\protag{2.4} {Definition}}
  Let $T$~be a smooth manifold.  A {\it family of M-theory data on closed
manifolds parametrized by~$T$\/} consists of:
 \roster
 \item"\rom(i\rom)" a family of closed spin Riemannian 11-manifolds $\Y\to T$
in the sense of \theprotag{1.1(i)} {Definition};
 \item"\rom(ii\rom)" a principal $E_8$-bundle $P\to \Y$ with a
connection~$A$; and
 \item"\rom(iii\rom)" a 3-form~$c\in \Omega ^3(\Y)$.
 \endroster  
        \enddefinition

\flushpar 
 Let $g$~denote the metric.  Associated to this data are
$\Cliff(\Y/T)$-modules $S\to \Y$ and $RS\to \Y$ generalizing those described
in the previous paragraph for a single manifold.  We also have a
$\Cliff(\Y/T)$-module $M(A)=S\otimes \Ad P$ constructed from the induced
connection on the real adjoint vector bundle; it also carries a compatible
quaternionic structure.

        \remark{\protag{2.5} {Remark}}
 A family of M-theory data is more information than a family of fields on
11-manifolds, all parametrized by~$T$.  For example, the connection~$A$ has
components in directions transverse to the fibers, as does the 3-form~$c$.
These extra components do not affect the definition of the
action~\thetag{2.6} below, but they do enter into the definition of the
covariant derivative on the line bundles~\thetag{2.2} and~\thetag{2.7}.  In
the bosonic functional integral one is meant to work once and for all with a
fixed family parametrized by equivalence classes of bosonic fields.
        \endremark

Assume that the fibers of~$\Y\to T$ are closed.  Then \thetag{2.3}~is, by
definition ~\cite{DFM,(4.12)},
  $$ \exp(-\Sgauge)\bigl(g,(A,c)\bigr) = \tau ^{1/2}_{D_{M(A)}}\cdot \tau
     ^{1/4}_{D_{RS}}\cdot \tau _{D_S}^{-3/4}\cdot \exp\Bigl(2\pi
     i\int_{\Y/T}\Omega \bigl(g,(A,c)\bigr) \Bigr). \tag{2.6} $$
Here $\tau ^{1/2}_{D_{M(A)}}$ is the global invariant~\thetag{1.29} for the
Clifford module~$M(A)$; the next two factors are defined in \theprotag{1.32}
{Remark}; and $\Omega \bigl(g,(A,c)\bigr)$~is an 11-form whose precise
formula is not of interest here.  Observe that the first and last factors
of~\thetag{2.6} are globally defined functions on~$T$, whereas the product of
the two middle factors is a section of a complex line bundle, namely the
bundle
  $$ L^{1/2}_{RS}\otimes \left( L_{S}^{1/2} \right)^{\otimes
     (-3)}\longrightarrow T \tag{2.7} $$
in the notation of~\S{1.3}. 

        \proclaim{\protag{2.8} {Theorem}}
 Let $\Y\to T$ be a family of M-theory data parametrized by~$T$ with closed
fibers.  Then the product $\exp(-\Sgrav)\cdot \exp(-\Sgauge)$ is a
well-defined function on~$T$. 
        \endproclaim

        \demo{Proof}
 Use the isomorphism of \theprotag{1.31} {Proposition} to construct a
trivialization of the tensor product of~\thetag{2.2} and~\thetag{2.7}.
        \enddemo

        \remark{\protag{2.9} {Remark}}
 The $E_8$~model for the $C$-field includes equivalences in the
groupoid~$\CY$ and it is crucial that \thetag{2.6}~is invariant under those
equivalences.  This is shown in~\cite{DFM,\S4}. 
        \endremark

        \remark{\protag{2.10} {Remark}}
 The setup of \theprotag{2.4} {Definition} and \theprotag{2.8} {Theorem}
includes the cancellation of gravitational anomalies.  To see this, let
$Y$~be a fixed closed spin 11-manifold and $\Met(Y)$ the space of metrics
on~$Y$.  Suppose $\Diff'(Y)$~is a group of spin diffeomorphisms of~$Y$ which
acts freely on the space of metrics.  Then
  $$ \bigl(\Met(Y)\times Y\bigr)/\Diff'(Y)\longrightarrow \Met(Y)/\Diff'(Y)
      $$
is a family of closed spin 11-manifolds with a canonical metric along the
fibers.  This is part of a family of M-theory data parametrized
by~$\Met(Y)/\Diff'(Y)$.  The application of \theprotag{2.8} {Theorem} to this
family is the statement that the quantum integrand is invariant
under~$\Diff'(Y)$.  
        \endremark

We interpret~\thetag{2.6} as an ``electric coupling'' factor in the
exponentiated action.  That is more evident in the heuristic
form~\thetag{2.3}, though the cubic self-coupling of the gauge field is a
notable departure from typical electric couplings.  The anomaly cancellation
and setting of the quantum integrand in \theprotag{2.8} {Theorem}---of a
fermionic pfaffian with an electric coupling---is an example of the
Green-Schwarz anomaly cancellation mechanism.

 \head
 \S{3} $(8k+3)$-Dimensional Manifolds with Boundary
 \endhead
 \comment
 lasteqno 3@ 22
 \endcomment

We return to the general mathematical theory of geometric invariants of Dirac
operators, now on compact manifolds with boundary.  There are two types of
boundary conditions for Dirac operators.  The global Atiyah-Patodi-Singer
boundary conditions~\cite{APS} are ubiquitous in the index theory literature.
Local boundary conditions do not exist for arbitrary Dirac operators in even
dimensions, so have not been as widely studied.  There is, however, a general
class of local boundary conditions---for both even and odd dimensions---which
lead to geometric invariants analogous to those in the closed case.  The
general story, in all dimensions, is discussed in Matthew Scholl's
thesis~\cite{Sch}.  Here we state refinements of his results in the
$(8k+3)$-dimensional case, as this is what we need for M-theory.

 \subhead \S{3.1} Generalities
 \endsubhead

Let $Y$~be an oriented odd dimensional Riemannian manifold with boundary and
$M\to Y$ a complex $\zt$-graded $\Cliff(Y)$-module.  Recall from~\S{1.2} that
there is an odd endomorphism~$\omega $ of~$M$---a multiple of Clifford
multiplication by the volume form---which is flat, commutes with the Dirac
operator~$D$, and squares to the identity.  Set $N=M\res{\bY}$.  We decompose
the restriction of~$\omega $ to~$\bY$ as
  $$ \omega =\gamma ^\nu \cdot \bom\qquad \text{on $\bY$}, \tag{3.1} $$
where $\gamma ^\nu $~is Clifford multiplication by the dual to the
outward-pointing unit normal vector, and so $\bom$~is Clifford multiplication
by a multiple of the volume form on~$\bY$.  Then $\gamma ^\nu $~and
$\bom$~are commuting endomorphisms of~$N$, with $\gamma ^\nu $~odd,
$\bom$~even, and
  $$ (\gamma ^\nu )^2 = (\bom)^2 = -\operatorname{id}_N.  $$
Decompose  
  $$ N = N_+ \oplus N_- \tag{3.2} $$
according to the eigenvalues~$\pm\sqmo$ of~$\bom$.  Since $\bom$~is even the
homogeneous pieces~$N^0,N^1$ decompose separately.  
 
The boundary~$\bY$ is an oriented Riemannian manifold and $N\to\bY$ restricts
to a $\Cliff(\bY)$-module.  Let $\gamma $~denote the Clifford action.  Then
$N$~with the $\zt$-grading~\thetag{3.2} is a $\Cliff(\bY)$-module for a
modified Clifford action: $\theta \in T_y^*(\bY)$ ~acts as~$\gamma ^\nu \gamma
(\theta )$.  The modified Clifford action preserves~$N^0\subset N$, whence  
  $$ N^0 = N^0_+\oplus N^0_- \tag{3.3} $$
is a $\zt$-graded $\Cliff(\bY)$-module.  
The associated Dirac operator is
  $$ \bD = \pmatrix 0&\bD_-\\\bD_+&0 \endpmatrix \tag{3.4} $$
relative to the decomposition~\thetag{3.2}.  By the previous remark
\thetag{3.4}~operates on sections of~\thetag{3.3}, i.e., maps sections
of~$N^0$ to sections of~$N^0$.

Assume now that $Y$~and $M$~are products near the boundary.  In other words,
postulate for some~$\epsilon >0$ an isometry of the product $(-\epsilon
,0]\times \bY$ onto a neighborhood~$U$ of~$\bY\subset Y$ and of the pullback
of $N\to\bY$ to $(-\epsilon ,0]\times \bY$ with~$M\res U$, including the
metric and covariant derivative.  Then $\gamma ^\nu $~extends over~$U$ as do
the decompositions~\thetag{3.1}, \thetag{3.2} and the Dirac
operator~\thetag{3.4}. On~$U$ the Dirac operator~$D$ decomposes as
  $$ D = \gamma ^\nu (\partial _\nu - \bD),  $$
where $\partial _\nu $~is differentiation along the unit normal vector,
defined using the product structure on~$M\res U$. 
 
Specialize to~$\dim Y=8k+3$ and $M$~quaternionic, as in~\S{1.3}.  The
quaternionic structure~$J$ commutes with~$\bom$, which has imaginary
eigenvalues, and since $J\:M^0\to \overline{M}^0$ it follows that
$J(N^0_{\pm})\subset \overline{N}^0_{\mp}$.  In particular, neither~$N^0_+$
nor~$N^0_-$ is quaternionic.  Rather, $J$~induces a pointwise symplectic
pairing
  $$ \lambda _y,\tlambda_y \longmapsto \langle J_y\lambda _y,\tlambda_y \rangle,\qquad
     \lambda _y,\tlambda_y\in N^0_y,\quad y\in \bY. \tag{3.5} $$
The homogeneous subspaces~$(N^0_+)_y$ and~$(N^0_-)_y$ are lagrangian and pair
nondegenerately under~\thetag{3.5}.  Assume $Y$~is compact.  By integration
we obtain a global version of the symplectic form:
  $$ \ll\! \lambda ,\tlambda \!\gg= \int_{\bY}\langle J\lambda ,\tlambda
     \rangle\;|dy|,\qquad \lambda ,\tlambda\in \Gamma
     _{\bY}(N^0). \tag{3.6} $$
The boundary Dirac operators
  $$ \bD_{\pm}\: \Gamma _{\bY}(N^0_{\pm})\longrightarrow \Gamma
     _{\bY}(N^0_\mp) \tag{3.7} $$
are skew-adjoint relative to~\thetag{3.6}.

Recall from~\thetag{1.21} that on~$Y$ we set $\Dirac=J\omega D\:\Gamma
_Y(M^0)\to\Gamma _Y(\overline{M}^0)$.  On a closed manifold this operator is
formally skew-adjoint, but on an arbitrary compact manifold the
skew-adjointness equation~\thetag{1.22} acquires a boundary term:
  $$ \int_{Y}\bigl\{ \langle {\Dirac\psi },\tpsi  \rangle + \langle
     \Dirac\tpsi  ,{\psi }\rangle \bigr\}\;|dy| = \int_{\bY}\langle
     J\bom\bpsi ,\btpsi  \rangle\;|dy|, \qquad \psi ,\tpsi \in \Gamma
     _Y(M^0). \tag{3.8} $$
Write $\psi \res{\bY} = \bpsi _+ + \bpsi _-$ and $\tpsi \res{\bY} = \btpsi _+
+ \btpsi _-$.  Then, up to a factor, the boundary term in~\thetag{3.8} may
be expressed in terms of~\thetag{3.6} as
  $$ \ll\! \bpsi _+,\btpsi_- \!\gg \;+\; \ll\! \btpsi _+,\bpsi_- \!\gg.
     \tag{3.9} $$
An elliptic boundary condition for~$\Dirac$ is a suitable
``half-dimensional'' subspace of~$\Gamma _{\bY}(N^0)$.  It determines a
formally skew-adjoint operator if the subspace is lagrangian with respect
to~\thetag{3.9}.  For a {\it local\/} elliptic boundary condition the
subspace is the space of sections of a lagrangian subbundle of~$N^0\to\bY$.

 \subhead \S{3.2} Global boundary conditions
 \endsubhead

Atiyah-Patodi-Singer~\cite{APS} introduced global boundary conditions for
first-order Dirac operators in the even dimensional case.  Generalizations
and the odd dimensional analog were studied in many works: see, for example,
\cite{SW} and the references therein.  We begin with a special class of
boundary conditions~\cite{J} used to construct the global
invariant~\thetag{3.11} below.
 
Let $\Y\to T$ be a family of compact $(8k+3)$-manifolds and $M\to\Y$ a
$\zt$-graded $\Cliff(\Y/T)$-module with compatible quaternionic structure.
Set~$N=M\res{\bYY}$.  On the boundary family $\partial \Y\to T$ there is an
induced $\zt$-graded $\Cliff(\bYY/T)$-module $N^0=N^0_+\oplus N^0_-$, as
in~\thetag{3.3} and associated family of Dirac operators~$\bD$, as
in~\thetag{3.4}.  Let $\bH=\bH_+\oplus \bH_-\to T$ be the bundle whose fiber
at~$t\in T$ is the space of $L^2$~sections of $N^0\res{\bYY_t}\to\bYY_t$.
Write $T=\bigcup U_a$ as in~\thetag{1.4} and recall from~\thetag{1.5} the
finite rank bundle $\bH(a)\to U_a$.  A global boundary condition for the
restriction of $\Y\to T$ over~$U_a$ is specified by a family of isometries 
  $$ I_t\:\bH_+(a)_t\to\bH_-(a)_t  $$
which are skew-adjoint relative to the pairing~\thetag{3.7}:
  $$ \ll\! I_t\lambda ,\tlambda \!\gg + \ll\! I_t\tlambda ,\lambda \!\gg
     =0,\qquad \lambda ,\tlambda\in \bH_+(a)_t.  $$
The corresponding boundary condition\footnote{In a Hamiltonian interpretation
\thetag{3.10}~projects out the negative energy modes.} on~$\psi \in \Gamma
_{\Y_t}(M^0)$ is
  $$ \bpsi _- \;+\; \left( I_t\oplus
     \frac{(\bD)_t}{\sqrt{(\bD)^2_t}} \right)\bpsi _+
     = 0, \tag{3.10} $$
where $\bpsi =\bpsi _+ + \bpsi _- \in \Gamma _{\bYY_t}(N^0_+\oplus N^0_-)$ is
the restriction of~$\psi $ to~$\bYY_t$ and we decompose ~$(\bH_+)_t$ into the
direct sum of~$\bH_+(a)_t$ and its orthogonal complement.  The Dirac
operator~$D_t^{(a,I)}$ is elliptic with the boundary conditions defined
by~$(a,I)$.  Moreover, it is also formally skew-adjoint: the boundary term
in~\thetag{3.8} vanishes for~$\psi ,\tpsi $ whose restriction to~$\bYY_t$
satisfies~\thetag{3.10}, as follows immediately from~\thetag{3.9} and the
skew-adjointness of the operator in~\thetag{3.10}.
 
The analytic properties of~$D_t^{(a,I)}$ are exactly the same as those
in the closed case~\cite{DF,Appendix~A}.  The geometric invariants of~\S{1.3}
are defined and \theprotag{1.26} {Proposition} holds.  We do not repeat them
here.   
 
There is an important generalization of this discussion which replaces the
subspace of boundary values defined by~\thetag{3.10} with a subspace~$W$
which is sufficiently ``close'' and is lagrangian with respect
to~\thetag{3.9}.  The space of admissible~$W$ forms a restricted
Grassmannian~\cite{Se2, Lecture~2}.  We refer to~\cite{Wo} for details.

The particular invariant $\tau ^{1/2}_{D^{(a,I)}}\:U_a\to \TT$ has a global
meaning when we consider its behavior under change of global boundary
condition~$(a,I)$.  Namely, it patches to a global section 
  $$ \tau ^{1/2}_D \:T\longrightarrow (\Pfaff {\bD })\inv
     \tag{3.11} $$
of the inverse pfaffian line bundle $(\Pfaff {\bD })\inv$ of the family of
Dirac operators on $\bYY\to T$ associated to the $\Cliff(\bYY/T)$-module
$N^0\to\bYY$.  In other words, a fiber of~$(\Pfaff {\bD })\inv$ is the line
of suitably equivariant functions~$\tau $ on the space of global boundary
conditions~$(a,I)$, or equivalently the line of suitably equivariant
functions on the restricted Grassmannian of subspaces~$W$.  The equivariance
is defined by a cocycle~$c$ which depends only on the boundary data:
  $$ \tau (a',I') = \tau (a,I)\,c\bigl((a',I'),(a,I) \bigr); \tag{3.12} $$
see~\cite{DF,Theorem~1.4} for details.  Note that~$|\tau ^{1/2}_D|=1$.  In
particular, $\Pfaff \bD $~is topologically trivial.  (It is not in general
geometrically trivial as $\nabla \tau _D^{1/2}$~may be nonzero.)

 \subhead \S{3.3} Local boundary conditions
 \endsubhead

Return now to a single manifold~$Y$, an $(8k+3)$-dimensional compact
Riemannian manifold; $M\to Y$ a quaternionic $\Cliff(Y)$-module; and
$N^0\to\bY$ the induced $\Cliff(\bY)$-module~\thetag{3.3}.

        \definition{\protag{3.13} {Definition}}
 A {\it local boundary condition\/} is a flat unitary section~$\epsilon $
of~$\End N^0\to\bY$ which is even relative to~\thetag{3.3}, satisfies
$\epsilon ^2=\operatorname{id}_{N^0}$, anticommutes with~$T^*(\bY)\subset
\Cliff(\bY)$, and anticommutes with~$J$ in the sense that~$J\epsilon
=-\bar{\epsilon }J$.
        \enddefinition

\flushpar
 The $\Cliff(\bY)$-action is defined after~\thetag{3.2}.  It follows that the
involutions~$\epsilon $ and $\bom$ ~commute, so we can separately
decompose~$N^0_+$ and~$N^0_-$ according to the eigenvalues~$\pm1$
of~$\epsilon $:
  $$ N^0_{\pm} = N^0_{\pm}[+]\oplus N^0_{\pm}[-].  $$
The domain
of the Dirac operator~$\Dirac^\epsilon $ with local boundary
condition~$\epsilon $ is the set of all~$\psi \in \Gamma _{Y}(M^0)$ whose
restriction~$\bpsi $ to the boundary satisfies
  $$ \epsilon (\bpsi)=\bpsi . \tag{3.14} $$
Because $\epsilon $~anticommutes with~$J$ the boundary term in~\thetag{3.8}
vanishes and $\Dirac^\epsilon $~is formally skew-adjoint.  Also, these local
boundary conditions are elliptic: the topological and geometric invariants
of~\S{1.3} are well-defined~\cite{Sch}.

Local boundary conditions always exist in odd dimensions.  For the
$(8k+3)$-dimensional case we have the special boundary conditions~$\epsilon
=\sqmo\,\bom$ and~$\epsilon =-\sqmo\,\bom$.  Furthermore, we can choose the
sign independently on different components of~$\bY$.

        \remark{\protag{3.15} {Remark}}
 Suppose $Y$~is spin and $M=S\otimes E\to Y$, where $S$~is the $\zt$-graded
spin bundle and $E$~is a real vector bundle.  Let $F=E\res{\bY}$.  On the
boundary the spin bundle splits into eigenspaces of~$\bom$, so for the
restriction of the even part we have $S^0\res{\bY}= \bS = \bS_+\oplus \bS_-$.
A local boundary condition ~amounts to a splitting~$F=F_+\oplus F_-$, and the
boundary condition~\thetag{3.14} translates to
  $$ \bpsi \in \Gamma _{\bY}(\bS_+\otimes F_+ \;\oplus \; \bS_-\otimes
     F_-). \tag{3.16} $$ 
If $\epsilon =-\sqmo\,\bom$ or~$\epsilon =\sqmo\,\bom$, then $F_-=0$
or~$F_+=0$ and \thetag{3.16}~ specializes to
  $$ \bpsi \in \Gamma _{\bY}(\bS_+\otimes F)\qquad \text{or}\qquad \bpsi \in
     \Gamma _{\bY}(\bS_-\otimes F).  $$
        \endremark

Since $\epsilon $~graded commutes with~$\Cliff(\bY)$, it anticommutes with
the Dirac operator~$\bD$, which therefore restricts to operators 
  $$ \bD[\pm]\:\Gamma _{\bY}\bigl(N^0_+[\pm] \bigr)\longrightarrow
     \Gamma _{\bY}\bigl(N^0_-[\mp] \bigr). \tag{3.17} $$
Also, as $\epsilon $~anticommutes with~$J$, the domain and codomain
in~\thetag{3.17} are dually paired (pointwise by~\thetag{3.5}) and the
operators in~\thetag{3.17} are formally skew-adjoint.  The operators
in~\thetag{3.17} are the Dirac operators~$\bD[+]$ and~$\bD[-]$
associated to the $\zt$-graded $\Cliff(\bY)$-modules
  $$ \aligned
      N^0[+] &= N^0_+[+] \oplus N^0_-[-] \\
      N^0[-] &= N^0_+[-] \oplus N^0_-[+]\endaligned  $$

Now suppose $T$~parametrizes a geometric family of Dirac operators on compact
$(8k+3)$-manifolds, as in the previous sections.  By contrast with the global
boundary conditions, a local boundary condition~$\epsilon $ may be defined
over the entire parameter space~$T$.  In particular, as remarked previously
the special choices~$\epsilon =\pm\sqmo\,\bom$ exist for all families.  

        \proclaim{\protag{3.18} {Theorem}}
 Let $D^\epsilon $~be a geometric family of Dirac operators on compact
$(8k+3)$-manifolds parametrized by~$T$ with local boundary
condition~$\epsilon $.  As in~\thetag{1.21} define $\Dirac^\epsilon =J\omega
D^\epsilon $.  Let $\bD[\pm]$~be the induced Dirac operators~\thetag{3.17} on
the boundary family.  Then there is a geometric isomorphism
  $$ \bigl(\Pfaff \Dirac^\epsilon \bigr)^{\otimes 2}\cong \bigl(\Pfaff
     \bD[+] \bigr)\otimes \bigl(\Pfaff \bD[-] \bigr)\inv
     . \tag{3.19} $$
        \endproclaim

\flushpar
 The isomorphism preserves the metric and covariant derivative.  There is no
compatible real structure; indeed, $\Pfaff \Dirac^\epsilon \to T$ may have
nontrivial curvature.  

While \thetag{3.19}~expresses the {\it square\/} of the pfaffian line bundle
explicitly, in general we cannot express~ $\Pfaff \Dirac^\epsilon $ directly
in terms of boundary data.  See, for example, \theprotag{1.31} {Proposition}
for the case where the boundary is empty.  But for the special boundary
conditions~$\epsilon =\pm\sqmo\,\bom$ we can say more.  For definiteness
take~$\epsilon =-\sqmo\,\bom$, so that the right hand side of~\thetag{3.19}
is~$\Pfaff \bD[+] = \Pfaff\bD$.  Recall from~\thetag{3.11} that there is a
global section~$\tau ^{-1/2}_D$ of this line bundle, with $|\tau
^{-1/2}_D|=1$.  The following simple observation defines a square root of
this bundle and section.

        \proclaim{\protag{3.20} {Lemma}}
 Let $K\to T$ be a smooth complex line bundle with metric and covariant
derivative, and $s\:T\to K$ a nonzero section.  Then there is a functorially
defined line bundle $K^{1/2}\to T$ with metric, covariant derivative, and
nonzero section $s^{1/2}\:T\to K^{1/2}$ together with an isomorphism
$(K^{1/2})^{\otimes 2}\cong K$ under which~$(s^{1/2})^{\otimes 2}=s$.  Also,
$|s^{1/2}|=|s|^{1/2}$ and $\nabla s^{1/2} = \frac 12 \frac{\nabla s}{s}
s^{1/2} $.
        \endproclaim

        \demo{Proof}
 The section~$s$ gives an isomorphism of~$K$ with the trivial bundle: the
geometry is characterized by the function~$|s|$ and the 1-form~$\frac{\nabla
s}{s}$.  Take~$K^{1/2}$ to be the trivial bundle with metric~$|s|^{1/2}$ and
covariant derivative~$\frac 12\frac{\nabla s}{s}$.  Under these isomorphisms
both the section~$s$ and its square root~$s^{1/2}$ are identified with the
constant function~$1$.
        \enddemo

        \proclaim{\protag{3.21} {Conjecture}}
 In the situation of \theprotag{3.18} {Theorem} for the special boundary
conditions~$\epsilon =\pm\sqmo\,\bom$ there are geometric isomorphisms 
  $$ \alignedat2
      \Pfaff \Dirac^\epsilon &\cong \bigl(\Pfaff \bD \bigr)^{1/2},\qquad
     &&\epsilon =-\sqmo\,\bom, \\
      \Pfaff \Dirac^\epsilon &\cong \bigl(\Pfaff \bD \bigr)^{-1/2},\qquad
     &&\epsilon =\phantom{-}\sqmo\,\bom.\endaligned \tag{3.22} $$
The square roots are defined by the sections~$\tau _D^{\mp1/2}$
of~$(\Pfaff\bD)^{\pm1}$ and \theprotag{3.20} {Lemma}.   More generally,
suppose $\bY = \sqcup(\bY)_i$ is written as a disjoint union of components
and on each~$(\bY)_i$ we choose either $\epsilon _i=\sqmo\,\bom$ or $\epsilon
_i=-\sqmo\,\bom$.  Then
  $$ \Pfaff \Dirac^\epsilon \cong \bigl(\Pfaff \bD \bigr)^{1/2}\otimes
     \bigotimes\limits_{{\text{$i$ such that}}\atop{\epsilon _i=\sqmo\,\bom}}
     \left(\Pfaff D^{(\bY)_i}\right)\inv ,  $$
where $\Dirac^{(\bY)_i}$~is the boundary Dirac operator~\thetag{3.4} on the
$i^{\text{th}}$~boundary component.
        \endproclaim

\flushpar
 Notice that if $\bY=\emptyset $ then \thetag{3.22}~reduces to
\theprotag{1.31} {Proposition}.

 \head
 \S{4} M-Theory Action on Compact Manifolds with Boundary
 \endhead
 \comment
 lasteqno 4@ 35
 \endcomment

The two types of mathematical boundary conditions discussed in~\S{3}---global
and local---correspond in physics to what we term {\it temporal\/} and {\it
spatial\/} boundaries.  In real time (Lorentzian signature) a temporal
boundary is a spacelike hypersurface.  In quantum field theory one associates
a Hilbert space to a temporal boundary, and then the functional integral
represents a state in the Hilbert space attached to the boundary.  Global
boundary conditions for fermionic fields determine a second state in the
Hilbert space, and the functional integral with fixed global boundary
conditions is the inner product of the two states.  A spatial boundary is
simply a boundary of space, so in the real time picture a boundary of each
spacelike hypersurface.  In quantum field theory spacetime locality requires
that boundary conditions for spatial boundaries be defined pointwise, so are
local.  They are part of the definition of the theory.

In this section we generalize \theprotag{2.8} {Theorem} to allow temporal or
spatial boundaries.  We first discuss the formal structure in general.

 \subhead \S{4.1} Actions and Anomalies
 \endsubhead

Consider a quantum field theory defined on a closed manifold~$Y$.  Let
$\BY$~denote the space of bosonic fields.  Fields with internal symmetry,
such as gauge fields and metrics, are best thought of as objects in a
groupoid, so we assume in general that $\BY$ is a groupoid.  The effective
exponentiated action, after integrating out any fermionic fields in the
theory, is a section
  $$ \exp({-\Seff})\:\BY\longrightarrow K_Y \tag{4.1} $$
of a line bundle\footnote{For our purposes a groupoid~$\scrB$ consists of a
manifold~$\scrB^0$ of objects, a manifold~$\scrB^1$ of morphisms, and a pair
of maps $\scrB^1\rightrightarrows \scrB^0$ which define the source and target
of a morphism.  (Of course there is more structure: an identity map,
composition laws, etc.)  A line bundle with covariant derivative on~$\scrB$
is a line bundle with covariant derivative $K\to\scrB^0$ and an isomorphism of
the two pullbacks of~$K$ to~$\scrB^1$.  (The isomorphism must satisfy
compatibilities which we do not spell out here.)}
 $K_Y\to\BY$ with metric and covariant derivative~$\nabla $.  The
notation in~\thetag{4.1} implies that morphisms (gauge transformations)
in~$\BY$ are lifted to~$K_Y$ and that $\exp(-\Seff)$~is invariant.  The
expression~$\exp({-\Seff})$ is meant to include everything: integrated out
fermionic fields, kinetic energy factors, Chern-Simons type factors, etc.  In
quantum field theory one imagines that the space~$\eBY$ of equivalence
classes of bosonic fields carries a measure and one defines the partition
function as the integral of~$\exp({-\Seff})$ over~$\eBY$ with respect to that
measure; correlation functions are defined similarly.  Even formally this
integral is not defined---one cannot add the values of~$\exp({-\Seff})$ at
different points as they lie in distinct lines.  Therefore, to define
formally the quantum theory we need as well a section
  $$ \triv\:\BY\to K_Y \tag{4.2} $$
which trivializes~$K_Y$ geometrically:
  $$ \aligned
      |\triv| &= 1 \\
      \nabla \triv&=0.\endaligned \tag{4.3} $$
Then  
  $$ \frac{\exp({-\Seff})}{\triv}\:\BY\longrightarrow \CC \tag{4.4} $$
is a global function that one could integrate over~$\eBY$ if one had a
measure.  In some theories the effective action is naturally a function, say
in a theory with only scalar fields and no fermionic fields, so a
trivialization~\thetag{4.2} does not enter explicitly, but in more
complicated theories the effective exponentiated action is naturally in the
form~\thetag{4.1} and a trivialization is needed.  The obstruction to the
existence of a trivialization~$\triv$ which satisfies~\thetag{4.3} is the
{\it anomaly\/}.  The factors~\thetag{2.1}, \thetag{2.6} in the effective
exponentiated M-theory action on a closed manifold each have this form.
\theprotag{2.8} {Theorem} map be interpreted as the construction of a
geometric trivialization~$\triv$ for their product.
 
As we mentioned in the introduction there can be different choices of
$\triv$, two choices differ by a locally constant function on $\BY$, and the
different phases on the connected components are moreover constrained by
locality conditions. Making a consistent choice is called {\it setting the
quantum integrand\/}.  See~\S{5.5} for a general discussion of the uniqueness
question.  Part of the significance of \theprotag{2.8} {Theorem} is that we
find a {\it canonical} isomorphism, allowing us to set the quantum integrand
for M-theory on closed 11-manifolds.

If $Y$~is compact with {\it spatial boundary\/} then the story is much the
same.  Here one needs local boundary conditions on all fields, which are part
of the definition of the space of fields.  With $\BY$~understood to be the
space (or groupoid) of bosonic fields satisfying the given local boundary
conditions, the discussion of the previous paragraph goes through unchanged.

Now suppose $Y$~is compact with {\it temporal boundary\/}.  Then there is a
space~$\bcY$ of allowed boundary conditions for all the fields; it is a fiber
bundle $\bcY\to\bcbY$ with base the space of boundary conditions~$\bcbY$ for
the bosonic fields.  The fibers are boundary conditions for the fermionic
fields, which are of global type.  In a well-defined theory one is meant to
do the functional integral for each~$f\in \bcY$ over the space of fields
whose boundary values equal~$f$.  In that case one obtains a ``function''
on~$\bcY$, which is allowed to be rather a section of a nontrivial line
bundle $\bK\to\bcY$.  In fact, in any theory we start with the classical
action and first integrate out the fermionic fields.  This yields the
effective exponentiated action, which is a section
  $$ \exp({-\Seff})\:\BYeff\longrightarrow K_Y \tag{4.5} $$
of a line bundle $K_Y\to\BYeff$.  Here $\BYeff$~is the fiber product
  $$ \CD
      \BYeff @>\pi >>\bcY\\
      @VVV @VV\rho V \\
      \BY @>>> \bcbY \endCD \tag{4.6} $$
(In other words, a field in~$\BYeff$ is a pair~$(b,f)$ consisting of a
bosonic field~$b$ on~$Y$ and a boundary condition~$f$ for both bosonic and
fermionic fields such that the boundary value of~$b$ is\footnote{For
groupoids of fields the fiber product is in the categorical sense: a field
in~$\BYeff$ includes a choice of isomorphism of the boundary value of~$b$
with the bosonic part of~$f$.} the bosonic part of~$f$.)  The bosonic
functional integral is an integral over the fibers of $\pi \:\BYeff\to\bcY$,
nominally with integrand~\thetag{4.5}.  In other words, we integrate over the
space of equivalence classes of bosonic fields with fixed boundary conditions
for both bosonic and fermionic fields.  But this is ill-defined as it stands,
even formally, as the integrand does not take values in a fixed line.  We
would like the line to depend only on the {\it boundary values} of the
bosonic fields on $Y$, since these are what are held fixed in the functional
integral. Therefore, to define the functional integral over the fibers
of~$\pi $ we must specify a generalization of the
trivialization~\thetag{4.2}: a line bundle
  $$ \bK\to\bcY \tag{4.7} $$
with metric and covariant derivative together with an isomorphism
  $$ \triv\:\pi ^*(\bK)\longrightarrow K_Y \tag{4.8} $$
which preserves the metric and covariant derivative.  The line bundle
$\bK\to\bcY$ is {\it not\/} derived from~\thetag{4.5}, but rather is
constructed prior to~\thetag{4.8}.  We explain this in~\S{5.4}.
Using~\thetag{4.8} we define
  $$ \triv\inv \circ \exp({-\Seff})\:\BYeff\longrightarrow \pi ^*(\bK),
     \tag{4.9} $$
a section of~$\pi ^*(\bK)\to\bcY$.  Its (formal) integral over the space of
equivalence classes in fibers of~$\pi $ is a section of $\bK\to\bcY$.  Note
\thetag{4.9} ~specializes to~\thetag{4.4} if~$\bY=\emptyset $. 
 
In~\S{5.4} we amplify this general discussion.  In particular, we relate this
Lagrangian (functional integral) point of view to the Hamiltonian approach to
anomalies. 

        \remark{\protag{4.10} {Remark}}
 To clarify the notation consider a theory with a gauge field~$A$ and a
spinor field~$\psi $ coupled to a vector bundle associated to~$A$.  Then
$\BY$~is the groupoid of gauge fields~$A$ on~$Y$ and the space~$\bcbY$ of
boundary conditions on the boson is the groupoid of gauge fields~$A^\partial
$ on~$\bY$.  For the fermion let $\H_{A^\partial }$~be the Hilbert space of
$L^2$~spinor fields~$\psi ^\partial $ on~$\bY$.  A boundary condition for the
spinor field~$\psi $ on~$Y$ is a ``half-dimensional'' subspace~$W\subset
\H_{A^\partial }$ which is roughly complementary to the boundary values of
harmonic spinor fields; cf. the discussion in~\S{3.2}.  So an object
in~$\bcY$ is a pair~$(A^\partial ,W)$ and an object in the groupoid~$\BYeff$
is a pair~$(A,W)$.  The partition function with boundary
condition~$(A^\partial ,W)$ is formally written as
  $$ \int\limits_{A\bigm|_{\bY}\cong A^\partial }\!\!\!dA\int\limits_{\psi
     \bigm|_{\bY}\in W}\!\!\!d\psi \;e^{-S(A,\psi )}.  $$
The result of the inner integral is the pfaffian of a Dirac operator, which
is~\thetag{4.5} in the general discussion above.  The outer integral is the
integral over equivalence classes in the fiber of~$\pi $ in~\thetag{4.6}.
        \endremark

 \subhead \S{4.2} Temporal Boundary Conditions
 \endsubhead

We resume our discussion of M-theory from~\S{2}, now on a compact
11-manifold~$Y$ with boundary, which in this subsection is assumed temporal.
The part of the effective exponentiated action which concerns us is the
product of the gravitino partition function~\thetag{2.1} and an electric
coupling ~\thetag{2.6} which depends on the $C$-field and the metric.  In the
notation of~\S{4.1} the groupoid of bosonic fields has objects
  $$ \BY = \bigl\{(g,C)\bigr\},  $$
where $g$~is a metric on~$Y$ and $C=(A,c)$ is a $C$-field, in the model
reviewed in~\S{2}.  The groupoid of boundary conditions on both bosonic and
fermionic fields has objects
  $$ \bcY = \Bigl\{ \bigl(\bg,\bC,W_{RS},W_S \bigr)\Bigr\},
      $$
where $\bg,\bC$~are the restrictions
 of~$g,C$ to~$\bY$ and $W_{RS},W_S$~are global boundary conditions~(\S{3.2})
for the Dirac operators~$D_{RS},D_S$ which appear in~\thetag{2.1}; the latter
are the boundary conditions on the fermionic fields.  The fiber
product~\thetag{4.6} is
  $$ \BYeff = \Bigl\{ \bigl(g,C,W_{RS},W_S \bigr)\Bigr\}.  $$
The generalization of \theprotag{2.4} {Definition} to compact manifolds with
boundary is the following.

        \definition{\protag{4.11} {Definition}}
   Let $T$~be a smooth manifold.  A {\it family of M-theory data on compact
manifolds with temporal boundary parametrized by~$T$\/} consists of:
 \roster
 \item"\rom(i\rom)" a family of compact spin Riemannian 11-manifolds $\Y\to T$
in the sense of \theprotag{1.1(i)} {Definition};
 \item"\rom(ii\rom)" a principal $E_8$-bundle $P\to \Y$ with a
connection~$A$; 
 \item"\rom(iii\rom)" a 3-form~$c\in \Omega ^3(\Y)$; 
 \item"\rom(iv\rom)" families of subspaces~$W_{RS},W_S$ which are global
boundary conditions for the operators~$\bD_{\bRS},\bD_{\bS}$.
 \endroster  
        \enddefinition

\flushpar 
 Here $S=S^0\oplus S^1\to \Y$ is the spinor bundle.  The induced boundary
$\Cliff(\bYY/T)$-module is $\bS = (S^0\res{\bYY})_+ \oplus
(S^0\res{\bYY})_-$; see~\thetag{3.3}.  For the Rarita-Schwinger
$\Cliff(\Y/T)$-module $RS = S\otimes T^*Y$ the induced
$\Cliff(\bYY/T)$-module is $\bRS \oplus \bS$, where
  $$ \bRS = \bS\otimes T^*(\bYY/T). \tag{4.12} $$

Now we are in a position to define the factors~\thetag{2.1}, \thetag{2.6} of
interest for a family of M-theory data with temporal boundary conditions.
For the gravitino this is straightforward: the Dirac
operators~$\Dirac_{RS},\Dirac_S$ with global boundary
conditions~$W_{RS},W_S$ determine pfaffian line bundles~\thetag{2.2} and a
section
  $$ \exp(-\Sgrav)\:T\longrightarrow \Pfaff\Dirac_{RS}\otimes
     (\Pfaff\Dirac_S)^{\otimes (-3)}. \tag{4.13} $$
For~\thetag{2.6} we note first that the last factor is simply a function
on~$T$.  The first factor is, according to~\thetag{3.11}, a global section
  $$ \tau ^{1/2}_{D_{M(A)}}\:T\longrightarrow (\Pfaff\bD_{N(\bA)})\inv 
      $$
of the pfaffian line bundle of the boundary family of Dirac operators on the
$\Cliff(\bYY/T)$-module $N(\bA)= \bS\otimes \Ad P\res{\bYY}$ with connection
induced from~$\bA=A\res{\bYY}$.  Now the product of the {\it square\/} of the
middle two factors is a global section of a product~$\scrL$ of pfaffian line
bundles defined from the boundary data (see~\thetag{3.11}):
  $$ \tau _{D_{RS}}^{1/2}\cdot \tau _{D_S}^{-3/2}\:T\longrightarrow
     \scrL:=\bigl(\Pfaff\bD_{\bRS\oplus \bS} \bigr)\inv \otimes
     \bigl(\Pfaff\bD_{\bS} \bigr)^{\otimes 3} . \tag{4.14} $$
\theprotag{3.20} {Lemma} determines a square root 
  $$ \tau _{D_{RS}}^{1/4}\cdot \tau _{D_S}^{-3/4}\:T\longrightarrow
     \scrL^{1/2} \tag{4.15} $$
for a hermitian line bundle with covariant derivative $\scrL^{1/2}\to T$
equipped with an isomorphism $(\scrL^{1/2})^{\otimes 2}\cong \scrL$.

        \proclaim{\protag{4.16} {Theorem}}
  Let $\Y\to T$ be a family of M-theory data parametrized by~$T$ with compact
fibers and temporal boundary.  Then there is a suitable
``trivialization''~$\triv$ such that $\bigl[\exp(-\Sgrav)\cdot
\exp(-\Sgauge)\bigr]/\triv$ is a section of a hermitian line bundle with
connection 
  $$ (\Pfaff\bD_{N(\bA)})\inv \otimes \scrK \longrightarrow T \tag{4.17} $$
which only depends on the boundary data.
        \endproclaim

\flushpar
 The bundle $\scrK\to T$ is defined in~\thetag{4.19} below.  According to the
discussion surrounding~\thetag{4.9}, the fact that \thetag{4.17}~depends only
on boundary data is what is needed to set up the functional integral.  We
remark that for fixed metric the line bundle~\thetag{4.17} was discussed
in~\cite{DFM,\S5}.  Also, \theprotag{4.16} {Theorem} should play an important
role in the extension of the computation of~\cite{DMW} to manifolds with
boundary.

        \demo{Proof}
 Let $\Ttil\to T$ be the fiber bundle whose fiber over~$t\in T$ is the
restricted Grassmannian of all possible boundary conditions~$W_{RS},\Wtil$
for the Dirac operators~$\Dirac_{RS},\Dirac_S$ at~$t$; see
\theprotag{4.11(iv)} {Definition}.  There is a section $s\:T\to\Ttil$ which
picks out the particular~$W_{RS},\Wtil$ in the family of M-theory data.  

By the definition of the line bundle in~\thetag{3.11} the product $
\tau^{1/2}_{D_{RS}} \tau^{-3/2}_{D_{S}}$ in~\thetag{4.14} lifts to an
equivariant complex-valued function on $\tilde T$; the equivariance condition
is~\thetag{3.12}.  The square root\footnote{This is a special case of
\theprotag{3.20} {Lemma} which we can make more explicit.  Let
$\widetilde{\CC^{\times }}\to\CC^{\times }$ be the double cover of the
nonzero complex numbers.  It carries a canonical function
$z^{1/2}\:\widetilde{\CC^{\times }}\to\CC$ which is equivariant for
the~$\zt$-action and squares to the identity on~$\CC^{\times }$.  It may be
viewed as a section of a hermitian line bundle on~$\CC^{\times }$ with
covariant derivative of order~2 whose square is isomorphic to the trivial
bundle.  In general we pull back from this universal case, for example here
by the map $\tau^{1/2}_{D_{RS}} \tau^{-3/2}_{D_{S}}\:\tilde T\to\CC^{\times
}$.} of this nonzero function is also equivariant and is a section of a
hermitian line bundle which we denote $\Ltil^{1/2} \to \tilde T$.  The
square~$(\Ltil^{1/2})^{\otimes 2}$ is canonically geometrically trivial.
Observe that a point in the fiber of $\scrL^{1/2}$ at~$t\in T$~is the space
of equivariant sections of~$\Ltil^{1/2}$ restricted to the fiber of~$\Ttil\to
T$ over~$t$.  Now for fixed boundary conditions~$(W_{RS},\Wtil)$ the argument
of \theprotag{1.31} {Proposition} applies to the Dirac
operators~$\Dirac_{RS},\Dirac_S$ to produce a trivialization
  $$ \triv\:T\longrightarrow \left[ \Pfaff\Dirac_{RS}\otimes
     (\Pfaff\Dirac_S)^{\otimes (-3)} \right]\otimes s^*\Ltil^{1/2}.
     \tag{4.18} $$
Dividing the product of~\thetag{4.13} and~\thetag{4.15} by~\thetag{4.18} we
obtain 
  $$ \frac{\exp(-\Sgrav)\cdot \tau _{D_{RS}}^{1/4}\cdot \tau
     _{D_S}^{-3/4}}{\triv}\:T\longrightarrow \scrL^{1/2}\otimes
     \bigl(s^*\Ltil^{1/2} \bigr)\inv .  $$
We claim that the line bundle 
  $$ \scrK := \scrL^{1/2}\otimes \bigl(s^*\Ltil^{1/2} \bigr)\inv
     \longrightarrow T \tag{4.19} $$
depends only on boundary data.  This is a nontrivial claim because each
factor, defined using \theprotag{3.20} {Lemma}, depends on the
$\tau^{1/2}$-invariant of the entire manifold with boundary.  Note first that
$\scrK^{\otimes 2}\cong \scrL$, since the square of~$\Ltil^{1/2}$ is
geometrically trivial.  Furthermore, as described above a section of~$\scrL$
is a suitably equivariant function on~$\Ttil$.  Hence there is an evaluation
map
  $$ \ev\:\scrL\longrightarrow s^*(\text{trivial}). \tag{4.20} $$
This may be regarded as a nonzero section of $s^*(\text{trivial})\otimes
\scrL^*\to T$, so by \theprotag{3.20} {Lemma} determines a square root, which
is in fact the inverse of the line bundle~$\scrK$ in~\thetag{4.19}.  As the
evaluation map~\thetag{4.20} depends only on boundary data, we are done.
        \enddemo

 \subhead \S{4.3} Spatial Boundary Conditions
 \endsubhead

As a preliminary we state the existence of a ``parity involution''~$\sigma $
on $C$-fields and its effect on the electric coupling~\thetag{2.6}.  We defer
the proof and discussion to~\S{5.1}.

        \proclaim{\protag{4.21} {Proposition}}
 Let $Y$~be a compact Riemannian spin 11-manifold with metric~$g$.  Then
there is an involution\footnote{This is to be understood in the categorical
sense: we are given an equivalence of ~$\sigma ^2$ with the identity
functor.}   $(A,c)\mapsto\sAc$ on the groupoid of $C$-fields such that
  $$ \exp(-\Sgauge)\bigl(g,\sAc\bigr) =
     \Bigl[\exp(-\Sgauge)\bigl(g,(A,c)\bigr)\Bigr]\inv . \tag{4.22} $$
        \endproclaim

\flushpar
 Recall that $\exp(-\Sgauge)\bigl(g,(A,c)\bigr)$ is an element of a complex
line, so implicit in~\thetag{4.22} is a functorial isomorphism between the
line for $\bigl(g,\sAc\bigr)$ and the inverse of the line for~$\bigl(g,(A,c)
\bigr)$.  The field strength~$G$ changes sign under~$\sigma $, and
\thetag{4.22}~is a refined version of the observation that \thetag{2.3}~is an
odd function of the local 3-form~$C$. This involution is relevant to the
parity-invariance of M-theory.   
 
In this paper we are interested instead in the induced involution on boundary
values of $C$-fields.  Recall the $\Cliff(\bY)$-modules~$\bS$ and~$\bRS$,
defined in and before~\thetag{4.12}.  Also recall that the Rarita-Schwinger
$\Cliff(Y)$-module $RS = S\otimes T^*Y$ induces the $\Cliff(\bY)$-module
$\bRS \oplus \bS$.  Finally, the $\Cliff(Y)$-module~ $M(A)=S\otimes \Ad P$
induces the $\Cliff(\bY)$-module $N(\bA)=\bS\otimes \Ad \bP$, where $A$~is a
connection on the $E_8$-bundle $P\to Y$ and $\bP\to\bY$ is its restriction to
the boundary.  Define the line
  $$ \multlinegap{45pt}\split \Lgauge(\bg,\bA,\bc) &= \bigl(\Pfaff
     \bD_{N(\bA)} \bigr)\inv \otimes \bigl(\Pfaff\bD_{\bRS\oplus \bS}
     \bigr)^{-1/2}\otimes \Bigl(\bigl(\Pfaff\bD_{\bS}
     \bigr)^{1/2}\Bigr)^{\otimes 3} \\
      &\cong \bigl(\Pfaff \bD_{N(\bA)} \bigr)\inv \otimes
     \bigl(\Pfaff\bD_{\bRS} \bigr)^{-1/2}\otimes \bigl(\Pfaff\bD_{\bS}
     \bigr).\endsplit \tag{4.23} $$
The square roots in the first line of~\thetag{4.23} are defined by the
functions~$\tau _{D_{RS}}^{1/2}$ and~$\tau _{D_S}^{-1/2}$ which play the role
of~`$s$' in \theprotag{3.20} {Lemma}; cf.~\thetag{3.11}.  The exponentiated
electric coupling~\thetag{2.6} lives in the line
  $$ \exp(-\Sgauge)\bigl(g,(A,c)\bigr)\in
     \Lgauge\bigl(\bg,(\bA,\bc)\bigr). \tag{4.24} $$

        \proclaim{\protag{4.25} {Proposition}}
 Continuing with \theprotag{4.21} {Proposition}, let $\sbAc$ be the
parity-reversal of the boundary values of~$(A,c)$.  Then there is a
functorial isomorphism
  $$ \multlinegap{45pt}\multline
      \Lgauge\bigl(\bg,\sbAc\bigr) \cong \Lgauge\bigl(\bg,(\bA,\bc)\bigr) \\
      \otimes \bigl(\Pfaff \bD_{N(\bA)} \bigr)^{\otimes 2} \otimes
     \bigl(\Pfaff\bD_{\bRS} \bigr)\otimes \bigl(\Pfaff\bD_{\bS}
     \bigr)^{\otimes (-2)}.  \endmultline \tag{4.26} $$
Furthermore, in a family this isomorphism preserves the metrics and covariant
derivatives.
        \endproclaim

Turning to M-theory on manifolds with spatial boundary, we note from the
beginning that in addition to the usual fields~$g,C,\psi $ there is a bosonic
field~$\Theta $ and a fermionic field~$\chi $ which live on the boundary.

        \definition{\protag{4.27} {Definition}}
    Let $T$~be a smooth manifold.  A {\it family of M-theory data on compact
manifolds with spatial boundary parametrized by~$T$\/} consists of:
 \roster
 \item"\rom(i\rom)" a family of compact spin Riemannian 11-manifolds $\Y\to T$
in the sense of \theprotag{1.1(i)} {Definition};
 \item"\rom(ii\rom)" a principal $E_8$-bundle $P\to \Y$ with a
connection~$A$; 
 \item"\rom(iii\rom)" a 3-form~$c\in \Omega ^3(\Y)$; 
 \item"\rom(iv\rom)" a principal $E_8$-bundle $Q\to\bYY$ with a
connection~$\Theta $; 
 \item"\rom(v\rom)" for each component~$(\bYY)_i$ of~$\bYY = \sqcup(\bYY)_i$
a choice $\epsilon _i=\sqmo\,\bom$ or~$\epsilon _i=-\sqmo\,\bom$; and
 \item"\rom(vi\rom)" a choice of boundary value isomorphism: see~\thetag{4.31}.
 \endroster  
        \enddefinition
 
We now examine the product of ~\thetag{2.1} and~\thetag{2.6} using local
boundary conditions based on the choice of~$\epsilon _i$.  The
restriction~$\bpsi$ of the Rarita-Schwinger field~$\psi \in \Gamma
_{\Y}(RS^0)$ to~$\bYY$ decomposes into tangential and normal components:
  $$ \bpsi = \bpsiT + \bpsin. \tag{4.28} $$
Let $(\bpsi)_i$ denote the restriction to~$(\bYY)_i$.  Then the boundary
condition for the gravitino operator~$D_{RS}$ is:
  $$ \left.{\bom(\bpsiT)_i = \pm\sqmo\,(\bpsiT)_i}\atop{\bom(\bpsin)_i =
     \mp\sqmo\,(\bpsin)_i}\right\} \qquad \text{as $\epsilon _i =
     \mp\sqmo\,\bom$}. \tag{4.29} $$
There are three ``ghost'' fields; see Appendix~A.  The boundary conditions
are given in~\thetag{A.12}, \thetag{A.13}, and \thetag{A.14}.  Those
in~\thetag{A.12} and~\thetag{A.13} lead to pfaffian line bundles which
cancel, due to the opposite signs in those equations.  The remaining
ghost~$\ghpsi$ has boundary condition~\thetag{A.14}, which we repeat here as
  $$ \bom(\bghpsi)_i = \mp\sqmo\,(\bghpsi)_i\qquad \text{as $\epsilon _i =
     \mp\sqmo\,\bom$}.   $$
Therefore, taking into account cancellations, \theprotag{3.21} {Conjecture}
implies that the effective exponentiated gravitino action~$\exp(-\Sgrav)$ is
a section of
  $$ \multlinegap{65pt}\multline
      \Lgrav=\left[(\Pfaff\bD_{\bRS})^{1/2} \otimes
     \bigl(\Pfaff\bD_{\bS}\bigr)^{-1}\right] \\
      \otimes \bigotimes\limits_{{\text{$i$ such that}}\atop{\epsilon
     _i=\sqmo\,\bom}} \left[ \bigl(\Pfaff D_{\bRS}^{(\bYY)_i} \bigr)\inv
     \otimes \bigl(\Pfaff D_{\bS}^{(\bYY)_i} \bigr)^{\otimes 2}
     \right]\longrightarrow T. \endmultline \tag{4.30} $$
Note that $\bpsin$ and $\ghpsi$~combine to give the second factor in each
bracketed expression on the right hand side of~\thetag{4.30}.

Let $\Theta _i$~be the restriction of~$\Theta $ to~$(\bYY)_i$.  The boundary
condition for the $C$-field~$(A,c)$ on~$(\bYY)_i$ specifies an isomorphism
with~$\Theta $ or its parity reversal, depending on the sign of~$\epsilon
_i$:
  $$ (\bA,\bc)_i \cong \cases (\Theta_i ,0) ,&\text{if $\epsilon
     _i=-\sqmo\,\bom$};\\(\Theta_i ,0)^\sigma ,&\text{if $\epsilon
     _i=\phantom{-}\sqmo\,\bom $.}\endcases \tag{4.31} $$
The isomorphism is a morphism in the groupoid of $C$-fields; the choice of
isomorphism is part of the data, and it enters~\thetag{4.32} below.
According to~\thetag{4.24} the exponentiated electric
coupling~$\exp(-\Sgauge)$ takes values in the line bundle~$\Lgauge$ defined
in~\thetag{4.23}.  The boundary condition~\thetag{4.31} and
isomorphism~\thetag{4.26} allow us to rewrite~$\Lgauge$ in terms of the
field~$\Theta $.  Let $N(\Theta _i)\to(\bYY)_i$ denote the $\Cliff(\bYY)_i
$-module $S^{(\bYY)_i}\otimes \Ad Q\res{(\bYY)_i}$.  Then the boundary
isomorphism~\thetag{4.31} determines an isomorphism of line bundles
  $$ \multline
      \Lgauge\cong \left[ \bigl(\Pfaff\bD_{\bRS} \bigr)^{-1/2}\otimes
     \bigl(\Pfaff\bD_{\bS} \bigr)\right] \otimes
     \bigotimes\limits_{{\text{$i$ such that}}\atop{\epsilon
     _i=-\sqmo\,\bom}} \bigl(\Pfaff D^{(\bYY)_i}_{N(\Theta _i)} \bigr)\inv \\
     \otimes \bigotimes\limits_{{\text{$i$ such that}}\atop{\epsilon
     _i=\sqmo\,\bom}} \left[ \bigl(\Pfaff D^{(\bYY)_i}_{N(\Theta _i)}
     \bigr)\otimes \bigl(\Pfaff D^{(\bYY)_i}_{\bRS} \bigr)\otimes
     \bigl(\Pfaff D^{(\bYY)_i}_{\bS}
     \bigr)^{\otimes(-2)}\right]\longrightarrow T.\endmultline \tag{4.32} $$
 
Finally, $\chi $ ~is a spinor field with values in~$\Ad Q$; the chirality
of~$\chi $ depends on the sign of~$\epsilon _i$.  Let $\chi _i$~be the
restriction of~$\chi $ to~$(\bYY)_i$.  Then 
  $$ \chi _i \in \cases N(\Theta _i)_+ ,&\text{if $\epsilon
     _i=-\sqmo\,\bom$};\\N(\Theta _i)_-,&\text{if $\epsilon
     _i=\phantom{-}\sqmo\,\bom$.}\endcases  $$
The exponentiated effective action for~$\chi _i$ is 
  $$ \exp(-S_{\chi _i}) = \pfaff \Dirac^{(\bYY)_i}_{N(\Theta _i)} \quad \text{
     or }\quad \bigl(\pfaff \Dirac^{(\bYY)_i}_{N(\Theta _i)} \bigr)\inv 
     \tag{4.33} $$
depending on the sign of~$\epsilon _i$, so the effective exponentiated action
for~$\chi $ is a section of
  $$ \Lchi=\bigotimes\limits_{{\text{$i$ such that}}\atop{\epsilon
     _i=-\sqmo\,\bom}} \bigl(\Pfaff D^{(\bYY)_i}_{N(\Theta _i)} \bigr)
     \;\;\;\otimes \; \bigotimes\limits_{{\text{$i$ such that}}\atop{\epsilon
     _i=\sqmo\,\bom}} \bigl(\Pfaff
     D^{(\bYY)_i}_{N(\Theta _i)}\bigr)\inv \longrightarrow T. \tag{4.34} $$
The tensor product of the factors of interest is therefore a section of the
tensor product 
  $$ \Lgrav\otimes \Lgauge\otimes \Lchi\to T  $$
of~\thetag{4.30}, \thetag{4.32}, and~\thetag{4.34}, which manifestly has a
geometric trivialization.  We summarize this discussion in the following,
which is rendered as a conjecture since it is based on \theprotag{3.21}
{Conjecture}.

        \proclaim{\protag{4.35} {Conjecture}}
   Let  $\Y\to T$  be  a family  of  M-theory data  parametrized by~$T$  with
compact fibers  and spatial  boundary.  Then the  product $\exp(-\Sgrav)\cdot
\exp(-\Sgauge)\cdot \exp(-S_\chi  )$ is a  section of a line  bundle over~$T$
which has a distinguished geometric trivialization~$\triv$.
        \endproclaim

\flushpar
 Therefore, the anomalies in the individual factors cancel and the theory is
well-defined with spatial boundary, at least from the point of view of
anomalies.  Notice that this anomaly cancellation and setting of the quantum
integrand work without restriction on the number of boundary components or
the topology of the fibers of~$\Y\to T$, as in Figure~1. 

 \midinsert
 \bigskip
 \centerline{
  \epsfxsize= 200pt
 \epsffile{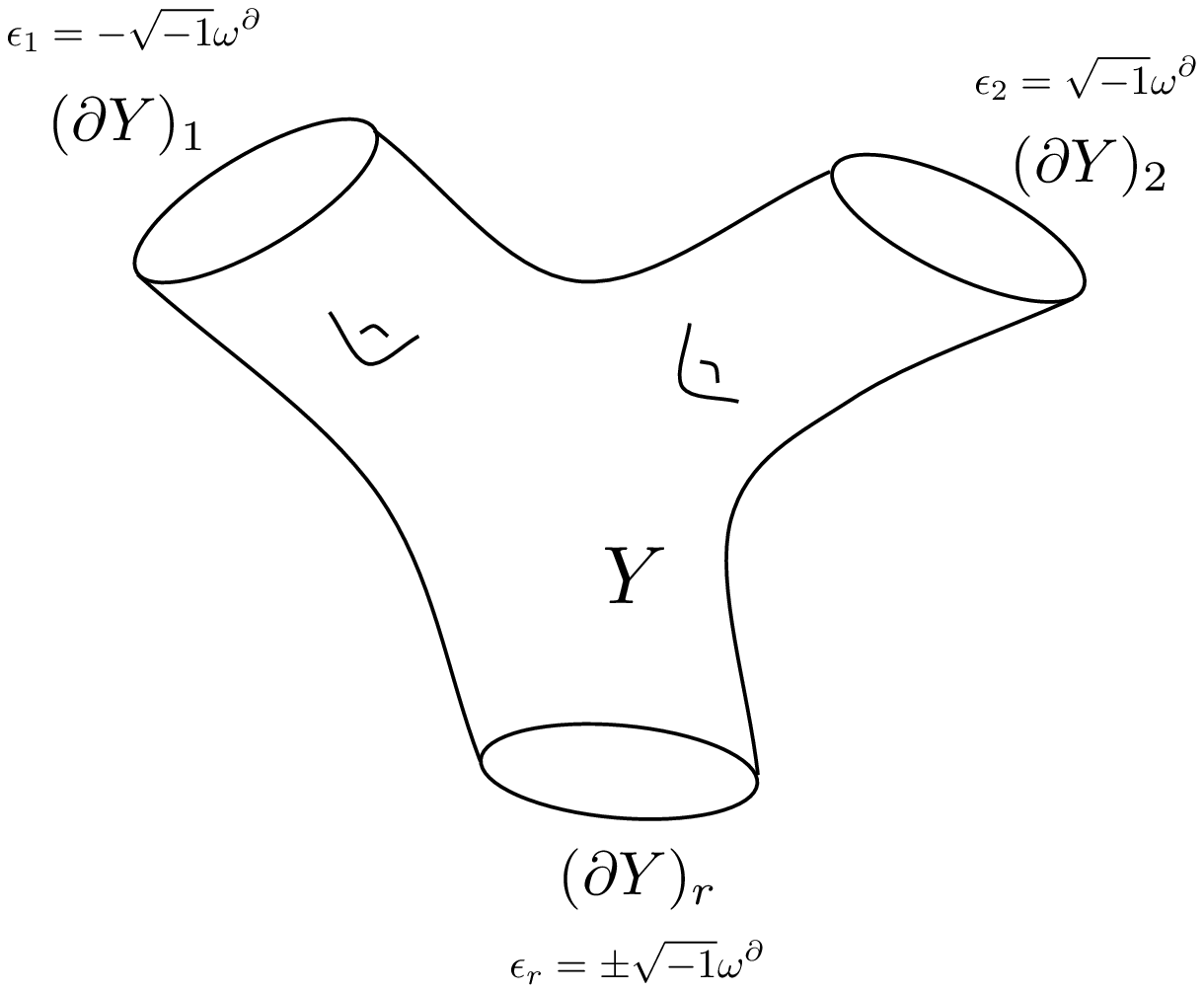}
}
 \nobreak
 \botcaption{Figure~1: M-theory is well-defined topologically on
11-manifolds with arbitrary numbers of spatial boundary components and
arbitrary chirality projects on each component. Each component carries an
independent $E_8$ super Yang-Mills multiplet.}
 \endcaption
 \bigskip
 \endinsert

 \head
 \S{5} Further discussion
 \endhead
 \comment
 lasteqno 5@ 13
 \endcomment

In this section we record brief remarks on several aspects of the main text.
We begin by outlining a construction of the involution (\theprotag{4.21}
{Proposition}) which is needed in \S{4.3}.  Next, in \S{5.2} we remark that
an analysis in 3~dimensions, which is parallel to that we did in
11~dimensions, sets the quantum integrand for an M2-brane.  M-theory also
admits M5-branes, and these are magnetically charged under the $C$-field.  We
do not know how to treat magnetic current with the $E_8$-model, however, so
we do not discuss M5-branes in the present paper.  The spatial boundary
condition for the $C$-field involves a choice of isomorphism~\thetag{4.31},
and in \S{5.3}~we explain the effect of this choice on the partition
function.  Finally, in~\S{5.4} we elaborate a bit on our discussion of
anomalies and setting the quantum integrand on manifolds with temporal
boundary.  In particular, we relate it to the Hamiltonian description of
anomalies, though we leave a full development for future investigation.

 \subhead \S{5.1} The $E_8$-model
 \endsubhead

The $E_8$-model for the gauge field~$C$, as introduced in~\cite{DFM} and
briefly reviewed in~\S{2} of this paper, has many nice features.  The fact
that the gauge action~\thetag{2.6} is expressed in terms of geometric index
theory invariants allows the direct definition of the partition function that
we have given in \theprotag{2.8} {Theorem}, \theprotag{4.16} {Theorem}, and
\theprotag{4.35} {Conjecture}.  Also, the connection with~$E_8$ is crucial
for understanding the Green-Schwarz anomaly cancellation for the $E_8\times
E_8$ Type~I supergravity.  However, this model has a defect: the algebraic
structure of $C$-fields is not at all explicit.  In other words, the space of
equivalence classes of $C$-fields is a torsor for an abelian
group,\footnote{It is not an abelian group because of the shift in the
quantization condition; see~\cite{W1}.} and so the groupoid of $C$-fields must
be a torsor in a categorical sense, i.e., a torsor for a Picard category.  We
do not need this structure, but in~\S{4.3} we do need an involution~$\sigma $
(compatible with the additive inverse on the associated abelian group).  We
indicate now how to construct~$\sigma $.
 
Quite generally, we can replace a groupoid with an {\it equivalent\/}
groupoid.

        \definition{\protag{5.1} {Definition}}
 An {\it equivalence\/} of groupoids~$\C_1,\C_2$ is a functor
$F\:\C_1\to\C_2$ such that there exists a functor $G\:\C_2\to\C_1$ and
natural transformations (homotopies) $G\circ F \Rightarrow \id_{\C_1}$ and
$F\circ G \Rightarrow \id_{\C_2}$.
        \enddefinition

\flushpar
 Gauge fields are modeled by groupoids, and equivalent groupoids are equally
good for the purposes of setting up the functional integral.  Of course,
different models have different advantages as we see here.  On the other
hand, we can use the equivalence to transport structure between equivalent
groupoids.  For example, suppose $\C_2$~in the definition has an
involution~$\sigma _2$, that is a functor $\sigma _2\:\C_2\to\C_2$ and a
natural transformation $\sigma _2\circ \sigma _2\Rightarrow \id_{\C_2}$.
Then $\sigma _1=G\circ \sigma _2\circ F$~is an involution on~$\C_1$. 
 
We apply this discussion to $C$-fields.  In~\cite{DFM,\S3.4} we introduce a
model for $C$-fields based on differential cocycles~\cite{HS} and prove that
it is equivalent to the $E_8$-model.  Let $\lambda (g)\in \Omega
^4(Y)$~denote the Pontrjagin form which represents half the first Pontrjagin
class.  We regard it as a singular cocycle with real coefficients.  Then an
object in the differential cocycle model is a triple~$(a, h, \omega )
\in C^4(Y; \ZZ) \times C^3(Y; \RR) \times \Omega ^4(Y)$ which satisfies 
  $$ \aligned
      \delta a&=0 \\
      \delta h&=\omega -a_{\RR} + \frac 12\lambda (g)\\
      d\omega &=0\endaligned  $$
The presence of~$\frac 12\lambda (g)$ reflects the shift in the quantization
law mentioned in the previous footnote.  The field strength of~$(a,h,\omega
)$ is~$\omega $.  To define an involution on these triples we need to use a
cocycle~$\lambda \in C^4(Y;\ZZ)$ which represents half the Pontrjagin class
as an integer cohomology class; see the next paragraph for a discussion.
Then the desired involution
  $$ (a,h,\omega )\longrightarrow (\lambda -a,-h,-\omega )  $$
sends the field strength to its opposite.  Transport it to the $E_8$-model to
define the involution~$\sigma $ of~\theprotag{4.21} {Proposition}.
 
We do not know a direct construction of the desired integer cocycle~$\lambda
$ on a Riemannian spin manifold~$Y$, i.e., a construction which only uses the
Riemannian metric.  We must introduce additional data; see~\cite{BryM} for
one such construction.  Our choice is to pass from the category of Riemannian
manifolds to an equivalent category whose objects are Riemannian spin
manifolds together with a classifying map of the principal spin bundle of
frames; the morphisms of the latter ignore the classifying map.  Then we fix
once and for all a cocycle~$\lambda _{\text{univ}}$ on the classifying space
and define~$\lambda $ as the pullback via the classifying map.

We must still prove~\thetag{4.22} and~\thetag{4.26}.  To do so it stands to
reason that the exponentiated action~$\exp(-\Sgauge)$ and the line
bundle~$\Lgauge$ should be defined in the model where the involution is
apparent; then we can check~\thetag{4.22} and~\thetag{4.26} in that model.
This is done in~\cite{FH}.  (In fact, that paper uses an equivalent model
closely related to the differential cocycle model.)

 \subhead \S{5.2} M2-branes
 \endsubhead

For our purposes here an M2-brane is a 3-dimensional compact
neat\footnote{`Neat' means that $\Sigma \cap\bY=\partial \Sigma $ and the
intersection is transverse.} spin submanifold~$\Sigma \subset Y$ of the
11-dimensional spin manifold of M-theory.  There is a fermionic field
on~$\Sigma $, a partner to the position of~$\Sigma $ in~$Y$, namely a spinor
field~$\lambda $ coupled to the half spin bundle of the normal bundle
of~$\Sigma $ in~$Y$.  Assume first that $\Sigma $~is closed.  We focus on two
factors in the exponentiated effective action.  The first is the effective
action~$\exp(-\Gamma _\lambda )$ of the spinor field~$\lambda $, the pfaffian
of a Dirac operator on~$\Sigma $.  This is analogous to~\thetag{2.1}.  The
second is analogous to~\thetag{2.3} and may be formally written as 
  $$ ``\vphantom{\Biggl(}\exp\left( 2\pi i \int_{\Sigma } C
     \right)\text{''}.  $$
As in~\thetag{2.6} we define it precisely in terms of the $E_8$-model for the
$C$-field:
  $$ \exp(-\SMgauge)\bigl(g,(A,c)\bigr) = CS_\Sigma (A)
     \cdot \tau _{D^\Sigma _S}^{-1/4}\cdot \exp\Bigl(2\pi
     i\int_{\Sigma }c \Bigr), \tag{5.2} $$
Here $CS_\Sigma $ is the exponentiated $E_8$~Chern-Simons invariant
(corresponding to the generator in~$H^4(BE_8;\ZZ)$) and the second factor is
a power of the $\tau $-invariant of the Dirac operator on~$\Sigma $ acting on
spinors~$S$.  Each of $\exp(-\Gamma _\lambda )$ and $\exp(-\SMgauge)$~is a
section of a line bundle with metric and covariant derivative.  The analog of
\theprotag{2.8} {Theorem}, which states that the product of these line
bundles has a canonical geometric trivialization, is proved with the same
analysis~(\S{1.3}) of Dirac operators in $8k+3$~dimensions, specifically
\theprotag{1.31} {Proposition}.
 
The extension to temporal boundaries for the M2-brane, in which
case~$\partial \Sigma \subset \bY$, proceeds analogously to~\S{4}.  We
comment only on the case of a spatial boundary, still assuming $\partial
\Sigma \subset \bY$. There is an additional factor involving fields on
$\partial \Sigma$, which plays a role analogous to ~\thetag{4.33}. This is
the partition function of a chiral, level 1, $E_8$~ WZW model coupled to the
$E_8$~gauge field~$\Theta$ restricted to $\partial \Sigma$. This factor
combines with the $E_8$ Chern-Simons term in~\thetag{5.2} to define the
membrane amplitude as a well-defined function.  These chiral degrees of
freedom have a fermionic formulation when the structure group of $\Theta$
reduces to $SO(16)$.  In this case the discussion is analogous to~\S{4.3}.

 \subhead \S{5.3} Boundary values of $C$-fields
 \endsubhead

For M-theory with spatial boundary there is a specified\footnote{That a
choice of isomorphism is needed was stated in~\S{4.1} in the words ``(4.6)~is
a fiber product in the categorical sense''.} isomorphism~\thetag{4.31} of the
boundary value of the $C$-field with the boundary $C$-field derived from the
$E_8$-connection~$\Theta $.  Let $\eBY(C)$~be the space of equivalence
classes of $C$-fields on~$Y$ and $\eBbY(\Theta )$ the space of equivalence
classes of $E_8$-connections on~$\bY$.  There is a subset $S\subset
\eBY(C)\times \eBbY(\Theta )$ of the product consisting of equivalence
classes which are isomorphic on~$\bY$ (with the boundary
condition~\thetag{4.31}).  The space of equivalence classes of fields in the
model with spatial boundary is a principal fiber bundle 
  $$ \eBY^{\text{eff}}(C,\Theta )\longrightarrow S \tag{5.3} $$
with structure group 
  $$ \prod\limits_iH^2\bigl((\bY)_i;\RZ \bigr). \tag{5.4} $$
The action of the structure group changes the isomorphism~\thetag{4.31} on
the boundary.  
 
The product 
  $$ \frac{\exp(-\Sgrav)\,\exp(-\Sgauge)\,\exp(-S_\chi
     )}{\triv}\:\eBY^{\text{eff}}(C,\Theta )\longrightarrow \CC  $$
is defined by \theprotag{4.35} {Conjecture}.  This function is {\it not\/}
constant on the fibers of~\thetag{5.3}, i.e., is not the pullback of a
function on~$S$.  Rather, $\lambda =(\lambda _i)\in \prod\limits_i
H^2\bigl((\bY)_i;\RZ \bigr)$ acts as multiplication
by~$\prod\limits_{i}\exp(2\pi \sqmo N_i\,\int_{(\bY)_i}\lambda _i)$ at a point
of~$\eBY^{\text{eff}}(C,\Theta )$, where $N_i$~is the index of the Dirac
operator~$D^{(\bY)_i}_{N(\Theta _i)}$.  There is a similar formula for the
effect on the partition function of an M2-brane $\Sigma \subset Y$; then
$N_i$~is replaced by~$\pm1$ depending on the sign of the boundary condition
$\epsilon _i=\pm\sqmo\,\bom$ in \theprotag{4.27(v)} {Definition}.
 
Consider the special case of heterotic M-theory~\cite{HW1}, \cite{HW2}.  Here
$Y=[0,1]\times X$ for a closed spin 10-manifold~$X$.  The signs in
\theprotag{4.27(v)} {Definition} are $\epsilon _0=-\sqmo\bom$ and $\epsilon
_1=\sqmo\bom$.  One recovers the usual heterotic fields on~$X$ by integrating
the M-theory fields over~$[0,1]$.  Thus the ``integral'' of~$\Theta $---the
sum over the two boundary components---is the $E_8\times E_8$~gauge field of
the heterotic theory and the integral\footnote{This integral may be defined
in the differential cocycle model discussed in~\S{5.1}; see~\cite{HS}.}
of~$C$ is the $B$-field of the heterotic theory.  This $B$-field is not
``closed'': its differential is computed from Stokes' theorem.  (For example,
the field strength $H\in \Omega ^3(Y)$ of~$B$ is a 3-form which satisfies
  $$ dH=\tr F_1^2 + \tr F_2^2 - \tr R^2,  $$
where $(F_1,F_2)$ is the curvature of the $E_8\times E_8$ bundle, $R$~is the
Riemannian\footnote{In a more precise treatment $R$~would be the curvature of
a connection with torsion determined by~$H$.} curvature, and the traces are
suitably normalized.)  Now the structure group~\thetag{5.4} is the product of
two copies of~$H^2(X;\RZ)$.  The diagonal acts trivially in the effective
heterotic theory.  For the anti-diagonal we identify~$H^2(X;\RZ)$ with the
group of isomorphism classes of flat gerbes.\footnote{These are the type of
``$B$-field'' which appear in Type~II theories.  They are ``closed''.}  Then
the action of the anti-diagonal adds an equivalence class of flat gerbes to
the given equivalence class of $B$-fields.  Said more precisely, the
$B$-field is a differential cochain of degree~3 which trivializes a fixed
differential cocycle of degree~4 constructed from the metric and $E_8\times
E_8$-connection.  The action in question adds a flat differential cocycle of
degree~3 to the $B$-field.  This can be measured through its effect on the
amplitude of a worldsheet instanton, realized as an open M2-brane in
heterotic M-theory.

 \subhead \S{5.4} Temporal boundaries and the Hamiltonian anomaly
 \endsubhead

The description in~\S{4.1} of anomalies in the case of temporal boundaries,
specifically the line bundle $\bK\to\bcY$, is related to the Hamiltonian
interpretation of anomalies.  The latter was explained by Fadeev and
Shatashvili ~\cite{Fa}, \cite{FS} and Segal~\cite{Se3} as follows.
Hamiltonian quantization of fermions on a manifold~$X$ leads to a bundle of
projective Hilbert spaces over the space of bosons on~$X$, and to quantize
the bosons one must lift this projective bundle to a vector bundle.  The
obstruction is measured topologically by an integral cohomology class of
degree three which, more importantly, has a natural geometric realization as
we now explain; see~\cite{MS}, \cite{CM}, \cite{CMM} for further discussion.
Set $X=\bY$ to be the boundary of a spacetime to make contact with~\S{4.1}.
Then this cohomology class is described by a line bundle with covariant
derivative $L\to\bcY[2]$ over the fiber product
  $$ \CD
      \bcY[2] @>p_2 >>\bcY\\
      @Vp_1VV @VV\rho V \\
      \bcY @>\rho >> \bcbY \endCD  $$
where $\rho $~is the map in~\thetag{4.6}.  Recall that the fiber of~$\rho $
is the space of fermionic boundary conditions for fixed bosonic boundary
conditions---a restricted Grassmannian of subspaces~$W$---and given two such
subspaces~$W,W'$ there is a natural determinant line~$L_{W,W'}$.  Further,
over the triple fiber product~$\bcY[3]$ there is an isomorphism
$L_{W,W''}\cong L_{W,W'}\otimes L_{W',W''}$.  This is the data of a gerbe
on~$\bcbY$.  The lift of the projective bundle to a vector bundle mentioned
above is encoded by a trivialization of this gerbe, which is a line bundle
with covariant derivative $\bK\to\bcY$ and an isomorphism 
  $$ p_1^*\bK\otimes L\longrightarrow p_2^*\bK \tag{5.5} $$
with appropriate compatibility on~$\bcY[2]$.  

The line bundle $\bK\to\bcY$ appears in~\thetag{4.7}.  Thus to define a
theory which includes temporal boundaries we must first resolve the
Hamiltonian anomaly on~$\bY$---i.e., produce the line bundle $\bK\to\bcY$
satisfying~\thetag{5.5}---and then specify the isomorphism~\thetag{4.8}.
This must be done in a consistent way for all manifolds~$Y$, constrained by
physical requirements such as locality (gluing laws).  

In the $8k+3$~dimensional situation considered in this paper, the
cancellation of anomalies (and setting of the quantum integrand) is
\theprotag{4.16} {Theorem}.  Furthermore, we expect that an analog of
\theprotag{1.31} {Proposition}, which we do not work out here, may be used to
construct $\bK\to\bcY$ directly from the boundary data.  Namely, the electric
coupling term determines a gerbe over~$\bcbY$, just as the Hamiltonian
quantization of fermions does, and the product of the two gerbes should have
a canonical trivialization.  (In this case both gerbes have a compatible real
structure.)  This is the Hamiltonian version of the Green-Schwarz mechanism.

It would be interesting to investigate further the many mathematical
questions underlying this brief sketch.

 \subhead \S{5.5} Topological terms
 \endsubhead

The trivialization~$\triv$ in~\thetag{4.2} which sets the quantum integrand
need not be unique.  The ratio of two choices is a locally constant function
  $$ Z\:\bcbY\longrightarrow \CC \tag{5.6} $$
of unit norm.  (We resume the notation of~\S{4.1}.)  Each
trivialization~$\triv$ must satisfy gluing laws, though we have not attempted
to make them precise here.  Nonetheless, there is a natural gluing assumption
for the ratio~\thetag{5.6} of two trivializations, namely that it has the
properties of the exponential of a {\it topological
term\/}~\cite{DeF,Chapter~6} in a classical action.  Thus it extends to
manifolds~$Y$ with boundary, it lies in a complex line which only depends on
the boundary data, and it is multiplicative under gluing.  Considering that
\thetag{5.6}~is also locally constant, we can simply say that it is the
partition function of a topological quantum field theory (TQFT) with
special properties.

        \definition{\protag{5.7} {Definition}}
 An {\it invertible topological quantum field theory (ITQFT)\/} on
oriented $n$-manifolds is an $n$-dimensional TQFT in which
 \roster
 \item"\rom(i\rom)" the complex vector space~$L_Z$ attached to an oriented
$(n-1)$-manifold~$Z$ is one-dimensional;
 \item"\rom(ii\rom)" there is a natural isomorphism $L_{\overline{Z}}\cong
L_Z\inv $, where $\overline{Z}$ is the oppositely oriented manifold; 
 \item"\rom(iii\rom)" the invariant $Z_Y$ of an $n$-manifold~$Y$ is nonzero;
and 
 \item"\rom(iv\rom)" $Z_{\overline{Y}}= Z_{Y}\inv $.
 \endroster  
An ITQFT is unitary if in addition $L_Z$~is hermitian and $|Z_Y|=1$.
        \enddefinition

\flushpar
 Note that in a general TQFT there is no analog of axioms~(ii) and~(iv); the
pairing between the vector space associated to~$Z$ and that associated
to~$\overline{Z}$ is induced from the cylinder on~$Z$; see~\cite{Qu},
\cite{Se2, Lecture~1}.  Also, recall that in a general unitary TQFT the
inverses in~(ii) and~(iv) are replaced by the complex conjugates.  The
category of manifolds on which the ITQFT is defined depends on the field
content.  For example, in M-theory it is the bordism category whose objects
are spin 10-manifolds equipped with a degree four integral cohomology class.
(We term this an ITQFT on 11-manifolds of the given type, as it is an
11-dimensional theory.)  We remark that a related notion appears
in~\cite{MZ}. 

Summarizing, ratios of settings of the quantum integrand are ITQFTs.
 
We do not have a definitive classification of ITQFTs in general, nor in the
case relevant to M-theory, but will report several observations.   

        \proclaim{\protag{5.8} {Proposition}}
 Let $Z$~be the partition function of an ITQFT on compact oriented
$n$-manifolds, and suppose $Z(S^n)=1$.  Then $Z$~is an oriented bordism
invariant of $n$-manifolds:  if $Y=\partial W$ for $W$~a compact oriented
$(n+1)$-manifold, then $Z(Y)=1$. 
        \endproclaim

        \demo{Proof}
  More generally, let $W$~be a bordism between closed oriented
$n$-manifolds~$Y_0$ and~$Y_1$.  Choose a Morse function $f\:W\to[0,1]$ with
$Y_0=f\inv (0)$, $Y_1=f\inv (1)$, and such that for each critical value~$c\in
(0,1)$ there is a unique critical point in~$Y_c=f\inv (c)$.  Then
for~$\epsilon >0$ sufficiently small, $Y_{c+\epsilon }$~is obtained
from~$Y_{c-\epsilon }$ by a surgery.  Namely, there is a compact
$n$-manifold~$Y_-$ with boundary diffeomorphic to $S^p\times S^q$, $p+q=n-1$,
so that 
  $$ \aligned
      Y_{c-\epsilon } &\simeq Y_- \cup (D^{p+1}\times S^q), \\
      Y_{c+\epsilon } &\simeq Y_- \cup (S^{p}\times D^{q+1}).\endaligned
      $$
Here $D^{r+1}$~is the $r$-dimensional ball with boundary~$S^r$.  The gluing
law implies that 
  $$ \aligned
      Z(Y_{c-\epsilon }) &= {Z(Y_-)}\inv \cdot Z(D^{p+1}\times S^q), \\
      Z(Y_{c+\epsilon }) &= {Z(Y_-)}\inv \cdot Z(S^{p}\times
     D^{q+1}),\endaligned \tag{5.9} $$
where the product is the inner product in the one-dimensional Hilbert space
attached to~$S^p\times S^q$.  Also, the gluing law applied to the
decomposition 
  $$ S^n = (D^{p+1}\times S^q)\cup(S^p\times D^{q+1})  $$
shows that 
  $$ 1=Z(S^n) = {Z(D^{p+1}\times S^q)}\inv \cdot Z(S^p\times
     D^{q+1}). \tag{5.10} $$
Combining~\thetag{5.9} and~\thetag{5.10} we conclude $Z(Y_{c-\epsilon
})=Z(Y_{c+\epsilon })$, whence $Z(Y_0)=Z(Y_1)$. 
        \enddemo

The converse, that the exponential of an $\RZ$-valued bordism invariant of
oriented $n$-manifolds defines an ITQFT with~$Z(S^n)=1$, is true.  The
argument, due to M. Hopkins, uses ideas from homotopy theory.  We do not give
it here. 
 
Next, observe that the product of two ITQFTs is again an ITQFT.  Now for
each~$\lambda \in \CC$ with~$\lambda \not= 0$ there is a simple ITQFT whose
partition function~$Z_\lambda $ on a closed $n$-manifold~$Y$ is
  $$ Z_\lambda (Y)=\lambda ^{\chi (Y)/2},  $$
where $\chi (Y)$~is the Euler characteristic.  In even dimensions we have
$Z_\lambda (S^n)=\lambda $, which proves the following. 

        \proclaim{\protag{5.11} {Corollary}}
  Let $Z$~be the partition function of an ITQFT on compact oriented
$n$-manifolds, $n$~even, and suppose $Z(S^n)=\lambda $.  Then $Z=Z_\lambda
\cdot Z'$, where $Z'$~is an oriented bordism invariant.
        \endproclaim

\theprotag{5.11} {Corollary} gives the complete story for oriented manifolds
in even dimensions.  We suspect that any ITQFT on compact oriented manifolds
in odd dimensions has $Z(S^n)=1$, thus is a bordism invariant, but we do not
have a proof in general.  For $n=3$~dimensions, the first nontrivial case,
there is a proof for oriented manifolds.

        \proclaim{\protag{5.12} {Proposition}}
  Let $Z$~be the partition function of an ITQFT on compact oriented
$3$-manifolds.  Then $Z(S^3)=1$. 
        \endproclaim

        \demo{Proof}
 Since the Hilbert space associated to~$S^2$ is one-dimensional, we have
$Z(S^1\times S^2)=1$.  Choose Heegaard decompositions 
  $$ \aligned
      S^1\times S^2 &\simeq Y_- \cup_{\Sigma _g} Y_+ \\
      S^3 &\simeq Y_- \cup_{\Sigma _g} Y_+ \endaligned \tag{5.13} $$
for $\Sigma _g$~a closed oriented 2-manifold of genus~$g\ge3$.  The gluing
maps in~\thetag{5.13} differ by an orientation-preserving diffeomorphism
of~$\Sigma _g$, and the ratio of~$Z(S^3)$ to~$Z(S^1\times S^2)=1$ is the
action of this diffeomorphism on the one-dimensional Hilbert space associated
to~$\Sigma _g$.  But that action is part of a one-dimensional representation
of the mapping class group, and for genus~$g\ge3$ the mapping class group is
perfect~\cite{P}, whence the action is trivial and $Z(S^3)=1$. 
        \enddemo

In higher dimensions we can show that if $n$~is odd, then $Z(S^n)$ is a
fourth root of unity.  For example, write 
  $$ \aligned
      S^5 &\simeq (D^3\times S^2)\cup_{S^2\times S^2}(S^2\times D^3) \\
      S^3\times S^2 &\simeq (D^3\times S^2)\cup_{S^2\times S^2}(S^2\times
     D^3) \endaligned  $$
and observe that the ratio of the gluing maps, the orientation-preserving
diffeomorphism $\left(\smallmatrix 0&1\\1&0  \endsmallmatrix\right)$
of~$S^2\times S^2$, has order~2.  Then 
  $$ \aligned
      S^3\times S^2 &\simeq \[(D^2\times S^1)\times S^2\]\cup_{(S^1\times
     S^1)\times S^2}\[(S^1\times D^2)\times S^2\] \\
      S^1\times S^2\times S^2 &\simeq \[(D^2\times S^1)\times
     S^2\]\cup_{(S^1\times S^1)\times S^2}\[(S^1\times D^2)\times S^2\] \\
      \endaligned  $$
and the ratio of the gluing maps is the orientation-preserving diffeomorphism
$\left(\smallmatrix 0&-1\\1&0 \endsmallmatrix\right)$ of~$S^1\times S^1$
times the identity of~$S^2$, which has order~4.  Finally, note that $Z(S^1
\times X) = 1$ for any $X$.  The argument continues to all higher
odd-dimensional spheres.
 
For the applications to quantum field theory, string theory, and M-theory we
need to complete these arguments and extend them to other bordism categories
of manifolds. 
 
Finally, we remark that there are ITQFTs in $n=3$~dimensions which
have~$Z(S^3)\not= 1$.  One such example is Chern-Simons theory for the
group~$U_1$ at the lowest level.  Then $Z(S^3)$~is a $24^{\text{th}}$~root of
unity.  This is not a contradiction to \theprotag{5.12} {Proposition}: recall
the ``framing anomaly'' of Chern-Simons theory~\cite{W2}.  It translates to the
assertion that Chern-Simons theory is defined on a different bordism
category.  For this particular case we also have a spin structure, so we
obtain the bordism category of {\it string\/} manifolds (or, in the older
literature, $MO\langle 8 \rangle$-manifolds).  The relevant bordism
group~$\Omega ^{\text{string}}_3$ is cyclic of order~24---it agrees with the
framed bordism group in this low dimension---and we conjecture that this
$U_1$ Chern-Simons theory is constructed from an isomorphism $\Omega
^{\text{string}}_3\to \ZZ/24\ZZ$.

 \head
 Appendix A: The Gravitino Path Integral
 \endhead
 \comment
 lasteqno A@ 14
 \endcomment

 \subhead {\S{A.1} The gravitino theory} 
 \endsubhead

In this appendix we review the formal definition of the partition function of
a gravitino as commonly used in supergravity.  (See~\cite{N}, \cite{K},
\cite{vN}, \cite{FT} for a sample of the literature. For a good recent
account of the boundary conditions in heterotic M-theory see~\cite{M}.)  As
we will note this subject has not been adequately investigated in the
literature, and consequently our discussion has some important gaps. Filling
these gaps is beyond the scope of this paper. In this appendix we simply
indicate how the standard discussion proceeds.

Let $Y$~be a Riemannian spin manifold of any dimension and $S\to Y$ a spin
bundle.  The gravitino field is a section of $S\otimes T^*Y$.  One may impose
chiral and/or Majorana projections.  The equations of motion are written
using an operator $R$ defined by composition of Clifford multiplication by
3-forms with the covariant derivative:
 
  $$ 
      R: \Gamma(S\otimes T^*Y) \to \Gamma(S\otimes T^*Y).  $$
In local coordinates $x^M$, $M=1,\dots, n$,  we have $ \psi = \psi_M dx^M $
and we define:   
  $$ 
      (R\psi)^M := \gamma^{MNP} \nabla_N \psi_P,
       $$
where $\gamma ^{MNP}$~denotes Clifford multiplication by the three-form
$dx^M\wedge dx^N\wedge dx^P$.  The equation of motion $R\psi =0$ follows from
the action
  $$ 
      \int_Y \bar \psi R \psi\; \vert dy \vert \tag{A.1} $$
We would therefore like to make sense of the formal gravitino path integral  
  $$ 
      Z = \int [d\psi] e^{- \int_Y \bar \psi R \psi \vert dy \vert}
      \tag{A.2} $$

In general, the path integral \thetag{A.2} formally vanishes, because the
action has an anticommuting gauge symmetry:
  $$ 
      \psi_M \to \psi_M + \nabla_M \epsilon
      \tag{A.3} $$
One may easily check that this is a local gauge symmetry of the action
\thetag{A.1} when the background Riemannian metric is Ricci-flat. In this case,
one attempts to define the true partition function $\CZ$ of the gravitino by
``dividing by the volume of the gauge group.'' Unfortunately, this volume is
zero, so one must proceed somewhat formally.

When the background metric is not Ricci-flat \thetag{A.3}~is not a local
symmetry of~\thetag{A.1}.  Nevertheless, the full supergravity action has a
local (super) gauge invariance which involves the
transformation~\thetag{A.3}.  However, if the background is not on-shell the
gauge transformations do not close into a super Lie algebra of symmetries and
one must use the BV formalism to quantize the theory.  This in turn leads to
substantial complications in the quantization of supergravity which we will
not discuss. Instead we will restrict attention in this appendix to
backgrounds with vanishing $G$-flux and Ricci-flat metrics.  This restriction
is a first significant gap in our discussion.

 \subhead {\S A.2 Local analysis of the equations of motion} 
 \endsubhead

Since the action \thetag{A.1} has a gauge symmetry, quantization involves
ghost fields.  In order to understand the nature of these ghosts let us
review briefly the standard discussion of the physical degrees of freedom of
the gravitino (see, for example, \cite{vN}, \cite{F3, Appendix}).  Here we
study the equations of motion in flat Minkowski space.  Thus $\SS=\SS^0\oplus
\SS^1$ is a $\zt$-graded real representation of the Lorentz spin group.  The
gravitino operator fits into a complex
  $$ 
      0 \to \Omega^0(\SS^0) ~{\buildrel \nabla \over \longrightarrow} ~
     \Omega^1(\SS^0) ~{\buildrel R \over \longrightarrow} ~ \Omega^1(\SS^1)
     ~{\buildrel \nabla^* \over \longrightarrow} ~ \Omega^0(\SS^1) ~{\buildrel
     \over \longrightarrow} ~ 0
      \tag{A.4} $$

The space of gauge inequivalent solutions of the equations of motion is the
degree one cohomology of~\thetag{A.4}.  In a flat Minkowski space of $n$
dimensions we analyze this cohomology as follows.  Write the wave equation in
a momentum basis
  $$ 
      \gamma^{MNP} k_N \psi_P =0
       $$
Now use
  $$  \gamma_M \gamma^{MNP} = (2-n) \gamma^{NP} $$
Hence, for $n-2\ne 0$,  
  $$ 
      \gamma^{MN}k_M \psi_N =0
       $$

Now use two more gamma matrix identities to write:  
  $$ 
     \gather
      \gamma^{MN} k_M \psi_N = (\gamma\cdot k)(\gamma \cdot \psi) + k \cdot
     \psi \tag{A.5}\\
      (\gamma^{MN} k_N)(\gamma\cdot \psi) = \gamma^{MNP}k_N \psi_P + \gamma^M
     (k \cdot \psi) - (\gamma\cdot k) \psi^M
      \tag{A.6} \endgather$$

From \thetag{A.5}, \thetag{A.6} we conclude that if we fix the gauge $\gamma
\cdot \psi =0$ then $k\cdot \psi =0$ and $(\gamma\cdot k) \psi^M =0$.  We
will henceforth fix this gauge. From $(\gamma\cdot k) \psi^M =0$ we learn
that for $\psi^M\ne 0$ we must have $k^2=0$.

Now, since $k^2=0$, any gravitino wave-function can be expanded  
  $$ 
      \psi_M = k_M s_1 + \bar k_M s_2 + \epsilon^{(i)}_M s_i
       $$
where $k^2 = \bar k^2 =0$; $k \cdot \bar k =1$; and $\epsilon^{(i)}$,
$i=3,\dots, n$, form a basis for the transverse space to the span of $k, \bar
k$. From $k\cdot \psi=0$ we learn $s_2=0$. Now we still have the gauge
freedom to shift $\psi_M \to \psi_M + k_M s_1$ preserving the gauge choice
$\gamma\cdot \psi=0$.  Thus, we can fix the gauge completely by taking
  $$ 
      \psi_M = \epsilon^{(i)}_M s_i
       $$
Note that we still have $\gamma^M \epsilon^{(i)}_M s_i =0$ and $(\gamma\cdot k) s_i  =0$, for each $i$. 

Now from $(\gamma\cdot k) s_i =0$ we learn the following. Choose a frame so
that $k= E(1, h,0^{n-2})$ with $h=\pm 1$.  Then $(\gamma\cdot k)= E \gamma^0
( 1 - h \gamma^0 \gamma^1) $ is proportional to a projection operator, so
under the decomposition
  $$ 
      Spin(1,n) \supset Spin(1,1) \times Spin(n-2)
       $$
we have $s_i \in [2^{[(n-2)/2]}]_h$ is an irreducible spinor of $Spin(n-2)$. 
Next from $\gamma^M \epsilon^{(i)}_M s_i =0$ we learn that $\epsilon^{(i)}_M s_i $ is in 
the irreducible representation of the tensor product of the vector of $Spin(n-2)$ with the 
spinor $[2^{[(n-2)/2]}]$. 

Thus, the cohomology gives the expected physical solutions. An important
lesson we may draw from this computation is that in the BRST quantization
there will be three ghosts. Two of these will be of the same chirality as the
gravitino, $k^M \psi_M$ and $k_M \epsilon$, while the third ghost,
$\gamma\cdot \psi$, will be of opposite chirality.

 \subhead \S A.3 {Partition function} 
 \endsubhead

We return to Euclidean field theory and so to a Riemannian manifold $Y$ with
$\zt$-graded spin bundle $S=S^0\oplus S^1\to Y$.

If we follow the paradigm of abelian gauge theory then the most natural
definition of the gravitino partition function follows from the
complex~\thetag{A.4}. Let $R^\perp$ be the restriction to $(\ker
R)^\perp$. Then the partition function should be
  $$ 
      \CZ = {\det R^\perp \det \nabla_M \over \det{}' (- \nabla^* \nabla)}
      \tag{A.7} $$
where $\nabla_M:\Omega^0(S^0) \to \ker R $.

Let us first formally justify~\thetag{A.7} using the BRST procedure. We fix
the gauge using
  $$ 
      \nabla^M \psi_M=s,  $$ 
where $s$ is an arbitrary spinor. This leaves unfixed the covariantly
constant spinors, which, by our assumption of Ricci-flatness, are the same as
the harmonic spinors.  We will deal with gauge fixing this finite dimensional
part of the gauge group below equation~\thetag{A.10}.  Following standard
procedure we now write:
  $$ 
      1 = \int_{\Omega^0(S^0)^\perp} d \epsilon\; \delta\bigl(s - \nabla^M
     (\psi_M+ \nabla_M \epsilon)\bigr) (\det{}' - \nabla^* \nabla)^{-1}
       $$
where $\Omega^0(S^0)^\perp$ is the orthogonal complement of $\ker \Dsl = \ker
(-\nabla^* \nabla) $.  This expression is to be inserted
in~\thetag{A.2}. Using gauge invariance of the action we may write:
  $$ 
      \CZ = \biggl( \int_{\Omega^0(S^0)^\perp} d \epsilon \biggr) \;
     \biggl(\int d\psi \delta\bigl(s - \nabla^M \psi_M \bigr) (\det{}' -
     \nabla^* \nabla)^{-1} e^{- \int \bar \psi R \psi \vert dy \vert}\biggr)
       $$
The formal division by the gauge group removes the first factor to produce
the partition function $\CZ$.  The remaining integral
 may be evaluated to give~\thetag{A.7}, by the choice of gauge $s=0$.

Unfortunately, \thetag{A.7}~is not in a form convenient for anomaly
analysis.  A second form follows from the formal BRST procedure by choosing a
different gauge, $s = \gamma \cdot \psi$, for an arbitrary spinor $s\in
\Omega^0(S^1)^\perp$.  We now write
  $$ 
      1 = \int_{\Omega^0(S^0)^\perp} d \epsilon \delta\bigl(s - \gamma^M
     (\psi_M+ \nabla_M \epsilon)\bigr) (\det{}' D^+)^{-1} 
      \tag{A.8} $$
for $D^+\:\Omega ^0(S^0)\to \Omega ^0(S^1)$ the Dirac operator.  We now
insert \thetag{A.8} into \thetag{A.2}, shift the field, and divide by the
volume of the gauge group to obtain the gauge-fixed expression
  $$ 
      \CZ = \int d\psi \delta( s- \gamma\cdot \psi) (\det{}'D^+)^{-1} e^{-
     \int \bar \psi R \psi} .
       $$
The ghost fields may be introduced by writing the determinant
in~\thetag{A.8} in terms of {\it commuting} ghost $\epsilon$ and antighost
$\beta$ fields as
  $$ 
      (\det{}'D^+)^{-1} = \int d\beta d \epsilon e^{-\int \bar \beta
     D^+ \epsilon} .
       $$

At this point, rather than setting $s=0$ we average over $s$ using the
expression  
  $$ 
      1 = {1\over (\det{}'D^-)^{1/2}} \int_{\Omega^0(S^1)^\perp} ds
     e^{-\int \bar s D^- s \vert dy \vert },
      \tag{A.9} $$
where $D^-\:\Omega ^0(S^1)\to \Omega ^0(S^0)$ is the Dirac operator.

We now invoke an algebraic identity. If $\phi_M = \psi_M + \half \gamma_M
(\gamma\cdot \psi)$ then
  $$ 
      -\bar \phi D^+_{T^*Y} \phi = \bar \psi R \psi - {1\over 4} (n-2)
       \overline{(\gamma\cdot \psi)} D (\gamma\cdot \psi) .
       $$

Thus, again formally dividing by the volume of the gauge group we  obtain  
  $$ 
      \CZ = {(\det D^+_{T^*Y})^{1/2} \over (\det{}' D^+)(\det{}'
     D^-)^{1/2} }
      \tag{A.10} $$
Since the path integral is independent of the choice of gauge we conclude
that \thetag{A.10}~is the same as~\thetag{A.7}.

There remains the gauge-fixing of the supersymmetry transformations by the
covariantly constant spinors. For this finite dimensional supergauge group we
will again insert 
  $$ 1 = \int_{\ker D} d \epsilon F(\epsilon, \psi , g) \tag{A.11} $$
where $F$ is a distribution concentrated on some gauge slice. (There does not
appear to be any particularly natural choice for $F$.)  Now the Berezin
measure $d\epsilon$ on the odd vector space $\ker D $ transforms as a section
of the line bundle $(\Det\ker D)^{-1}$.  The latter is identified
with~$\Pfaff\Dirac$.  By~\thetag{A.11} $F(\epsilon,\psi,g)$ transforms in the
inverse line bundle. In the gauge fixing procedure we factor out the gauge
group leaving behind $F(0,\psi , g)$ in the path integral.  The product
$F(0,\psi_*(g),g) \cdot \CZ$ is a section of a determinant line bundle.
Specifically, for $\dim Y=11$ we identify~$S^0$ and~$S^1$ using the volume
form~(\S{1.2}) and rewrite~\thetag{A.10} using pfaffians~(\S{1.3}).  This
yields~\thetag{2.1} and provides the motivation for our choice of line bundle
with covariant derivative in~\thetag{2.2}.

 \subhead \S A.4 {Boundary conditions for ghosts}
 \endsubhead

Let us finally consider the analysis of the gravitino determinant for $\dim
Y=11$ in the presence of a boundary, as in~\S{4.3}.  For further simplicity
we will assume the boundary Dirac operator $D^{\p} $ has no zeromodes. The
boundary conditions on the gravitino are given in~\thetag{4.29}, and we make
a definite choice of sign.  Recall that the restriction~$\bpsi$ of the
Rarita-Schwinger field~$\psi $ to~$\bY$ decomposes into its tangential and
normal components~\thetag{4.28}.  Then the boundary conditions are:
  $$ 
     \aligned
      \bom \bpsiT &= + \sqmo\,\bpsiT \\ 
      \bom \bpsin &= - \sqmo\, \bpsin  \endaligned
       $$
where $T$ denotes the tangential component and $\nu$ the normal component.
These boundary conditions imply that the gauge group must be restricted by
  $$ 
      \eqalign{ \bom \nabla_T \epsilon^\partial & = + \sqmo\,\nabla_T
     \epsilon^\partial \cr
      \bom \nabla_\nu \epsilon^\partial & = -
     \sqmo\,\nabla_\nu\epsilon^\partial \cr}
      \tag{A.12} $$
We then choose boundary conditions on $\beta$ so that $D^+$ is
skew-adjoint:
  $$ 
      \eqalign{ \bom \nabla_T \beta^\partial & = -\sqmo\,\nabla_T
     \beta^\partial \cr
      \bom \nabla_\nu \beta^\partial & = +
     \sqmo\,\nabla_\nu\beta^\partial \cr}
       \tag{A.13} $$
The third ghost determinant comes from integrating over~$s$ \thetag{A.9},
which has the same boundary condition as~$\gamma \cdot \psi $: 
  $$ \bom(s^\partial ) = -\sqmo\,s^\partial . \tag{A.14} $$

The boundary conditions ~\thetag{A.12} and~\thetag{A.13} do not fit the
discussion on local boundary conditions for Dirac operators given in~\S{3.3}.
We can relate them to the standard ones, at least topologically.  Note first
that if there are no covariantly constant spinors on the boundary, which we
assume, then the first equation in~\thetag{A.12} is equivalent to the
condition $\epsilon^\p \in \Omega ^0_{\bY}(S^0_+)$.  Similarly, the first
equation in~\thetag{A.13} is equivalent to the condition $\beta^\p \in \Omega
^0_{\bY}(S^1_-)$.  Let us define
  $$ 
      \widetilde\Omega^0(S^0) := \{ \epsilon\in \Omega^0(S^0) | \epsilon^\p
     \in \Omega ^0_{\bY}(S^{0}_+), \;(\nabla_\nu \epsilon)^\p \in \Omega
     ^0_{\bY}(S^0_-) \}.  $$
We then have the diagram

\input diagrams

\diagram
0 & \rTo & \widetilde \Omega^0(S^0) & \rTo & \Omega^0(S^0)_{\epsilon^\p \in \Omega ^0_{\bY}(S^0_+)} & \rTo^{\pi_+\circ\nabla_\nu} & \Omega^0(S^{0,\p}_{+}) & \rTo & 0 \\
  &      & \dTo^{D^+}              &      & \dTo^{D^+}                           &                             & \dTo^{\gamma^\nu}      &      &  \\
0 & \rTo & \Omega^0(S^1)_{\beta^\p \in \Omega ^0_{\bY}(S^0_-)} & \rTo & \Omega^0(S^1)  & \rTo^{\pi_+\circ\iota^*} & \Omega^0(S^{1,\p}_{+}) & \rTo & 0 \\
\enddiagram
 where $\pi_+$ is the projection to the positive chirality spinors and
$\iota^*$ is pullback to the boundary.  Each vertical arrow is a Fredholm
operator.  The right-most arrow is an isomorphism, and hence it follows that
topologically the determinant lines (and Pfaffian lines) associated to the
other two vertical arrows are isomorphic. We expect that the metrics and
covariant derivatives similarly match up, but we have not checked the
details.

 \head
 Appendix B: Quaternionic Fredholms and Pfaffians
 \endhead
 \comment
 lasteqno B@  3
 \endcomment

We prove a theorem about a special space of Fredholm operators which is a
topological version of \theprotag{1.31} {Proposition}.  We include it here to
emphasize the topological nature of the latter.  We remark that the similar
\theprotag{1.16} {Proposition} is not topological; there is no analog for
Fredholm operators.\footnote{The space of self-adjoint Fredholm operators on
a complex Hilbert space has only one natural real line bundle, analogous
to~\thetag{1.15} and the construction below; there is no nontrivial (real)
determinant line bundle.}  For determinants and pfaffians on spaces of
Fredholm operators, see~\cite{Q} and~\cite{Se1}. 
 
Dirac operators on compact manifolds are special examples of Fredholm
operators.  Whereas Dirac operators have discrete spectrum with no
accumulation points, a basic fact for the geometric constructions of~\S{1},
general Fredholm operators have continuous spectrum.  Spaces of Fredholm
operators are classifying spaces for $K$-theory~\cite{AS}.

Let~$\H$ be a separable complex Hilbert space with a unitary quaternionic
structure $J\:\H\to\Hb$.  Thus $\Jb J=-\id_{\H}$ and $J$~is skew-adjoint in
the sense that
  $$ \langle Jv,w \rangle + \langle Jw,v \rangle =0,\qquad v,w\in
     \H.  $$
Define 
  $$ \F = \{T\:\H\to\Hb: \text{$T$ is Fredholm}, \; \text{$T$ is
     skew-adjoint}, \; \Tb J = \Jb T\}.  $$
This space of skew-adjoint quaternion Fredholm operators has two contractible
components and a third component which is a classifying space for~$KR^{-3}$.
There are two natural real line bundles $\Pfaff\to\F$ and~$L^{1/2}\to\F$
which we now construct.

A subspace~$A\subset \H$ is said to be {\it quaternionic\/} if~$J(A)=\Ab$.
For finite dimensional quaternionic~$A\subset \H$ we define the open set 
  $$ U_A = \{T\in \F : T(A)\subseteq \Ab, \; T(\H)\oplus \Ab =
     \Hb\}.  $$
Let $\Pfaff_A\to U_A$ be the real line bundle whose fiber at~$T\in U_A$ is
the pfaffian of the skew-adjoint operator $T\res A\:A\to\Ab$.  The real
structure is given by~$\pfaff J\res A$.  (See the paragraph
containing~\thetag{1.23} for a discussion of finite-dimensional pfaffians.)
If $A\subset B$ then on~$U_A\cap U_B$ there is a canonical isomorphism 
  $$ \Pfaff_B \cong \Pfaff_A\otimes \Pfaff_{B/A},  $$
where the fiber of~$\Pfaff_{B/A}$ at~$T$ is the Pfaffian line of $T\:B/A \to
\overline{B/A}$.  The pfaffian of that operator is nonzero and real, and this
gives an isomorphism $\Pfaff_A\to\Pfaff_B$.  For $A\subset B\subset C$ a
cocycle identity is satisfied, so by patching we obtain a global real line
bundle $\Pfaff\to \F$.

To construct the bundle $\Lh\to\F$ we observe that for~$T\in \F$ the
composition $S^{(T)}=\Jb T\:\H\to\H$ is self-adjoint Fredholm and quaternion
linear in the sense that~$JS^{(T)} = \overline{S^{(T)}}J$.  So its
eigenspaces are quaternionic.  For $A\subset \H$ finite dimensional
quaternionic set $S^{(T)}_A=S^{(T)}\res A$.  Let $\Lh_A\to U_A$ be the real
line bundle whose fiber at~$T$ is
  $$ \left\{f_A \:\RR\setminus \HH\spec(S^{(T)}_A) \longrightarrow \RR \;:\;
     f_A (\beta ) \;=\; (-1)^{\#\left\{{{\alpha <\lambda<\beta }\atop{\lambda
     \in \HH\spec(S^{(T)}_A)}}\right\}}\;f_A (\alpha ) \right\}. \tag{B.1} $$
On~$A\subset B$ we note that $S^{(T)}_{B/A}\:B/A\to B/A$ is invertible, so the
spectrums of~$S^{(T)}_A$ and~$S^{(T)}_B$ agree in a neighborhood of~$0\in \RR$.
Define an isomorphism $\Lh_B\to\Lh_A$ which identifies~$f_A$ and~$f_B$ if
$f_A=f_B$~in this neighborhood of zero.  This satisfies a cocycle identity,
so defines a global line bundle $\Lh\to\F$.  It carries a natural metric.

        \proclaim{\protag{B.2} {Proposition}}
 The real line bundles~$\Pfaff\to\F$ and~$\Lh\to\F$ are isomorphic. 
        \endproclaim

\flushpar
 Equivalently, the tensor product $\Lh\otimes \Pfaff\to\F$ is trivializable.
For this topological theorem we make a choice (of a norm) to specify a
trivialization.  In the result for Dirac operators, \theprotag{1.31}
{Proposition}, the trivialization is required to have unit norm, and this
leads to the canonical construction in the proof of that result.

        \demo{Proof}
 The space~$\F$ is paracompact, so use a partition of unity to construct a
metric on~$\Pfaff\to\F$.  Set $J_A=J\res A$.  The bundle $\Pfaff_A\to U_A$
has a trivialization~$\pfaff J_A$.  Under the patching isomorphism
on~$U_A\cap U_B$ the ratio of the trivializations of~$\Pfaff_B$
and~$\Pfaff_A$ is the pfaffian of the self-adjoint operator~$S^{(T)}_{B/A}\:B/A\to
B/A$, that is, the product of its (real) eigenvalues viewing $S^{(T)}_A$~ as a
quaternion linear operator.  Use instead the unit norm trivialization $\pfaff
J_A/|\pfaff J_A|$ on~$U_A$.  Then the transition function on~$U_A\cap U_B$
is
  $$ \frac{\pfaff S^{(T)}_{B/A}}{|\pfaff S^{(T)}_{B/A}|}. \tag{B.3} $$
Since $A$~is finite dimensional the spectrum of~$S^{(T)}_A$ is finite, so the
domain of~$f_A $ in~\thetag{B.1} contains an interval~$(-\infty ,\alpha )$
for some~$\alpha \in \RR$.  Let $f_A (-\infty )$~denote the value of~$f_A$ on
this interval.  Trivialize~$\Lh_ A\to U_A$ by the function~$f_A$
with~$f_A(-\infty )=1$.  The transition functions for this trivialization are
exactly~\thetag{B.3}.
        \enddemo

\NoRunningHeads
\widestnumber\key{SSSSSSS}   

\enddocument